\documentclass{aa} 
\usepackage{graphicx}
\usepackage[super]{nth}
\usepackage{txfonts}
\usepackage{natbib}
\usepackage{mathtools}
\usepackage{breqn}
\usepackage{float}
\usepackage{amsmath}
\usepackage{esint}
\usepackage{subfig}

\usepackage{multirow}
\usepackage[percent]{overpic}
\usepackage{xcolor}
\graphicspath{{./figures/}}
\bibpunct{(}{)}{;}{a}{}{,} 
\begin{document}

        \title{Study of the propagation, in situ signatures, and geoeffectiveness of shear-induced coronal mass ejections in different solar winds}
        
        \titlerunning{Study of the propagation, in situ signatures, and geoeffectiveness of shear-induced coronal mass ejections}        
        
        \author{D.-C. Talpeanu \inst{1,2}
                \and
                S. Poedts \inst{1,3}
                \and
                E. D'Huys\inst{2}
                \and
                M. Mierla \inst{2,4}
        }
        
        \institute{Centre for mathematical Plasma Astrophysics (CmPA), Department of Mathematics, KU Leuven, Celestijnenlaan 200B, 3001 Leuven, Belgium\\
                \email{dana.talpeanu@observatory.be}
                \and
                SIDC - Royal Observatory of Belgium (ROB), Av. Circulaire 3, 1180 Brussels, Belgium               
                \and
                Institute of Physics, University of Maria Curie-Sk{\l}odowska, PL-20-031 Lublin, Poland
                \and 
                Institute of Geodynamics of the Romanian Academy, Jean-Louis Calderon 19-21, 020032 Bucharest, Romania
        }
        
        \date{Received : to be filled in; accepted: to be filled in}
        
        
        \abstract
        {}
        {Our goal is to propagate multiple eruptions ---obtained through numerical simulations performed in a previous study---  to $1\;$AU and to analyse the effects of different background solar winds on their dynamics and structure at Earth. We also aim to improve the understanding of why some consecutive eruptions do not result in the expected geoeffectiveness, and how a secondary coronal mass ejection (CME) can affect the configuration of the preceding one.}
        {Using the 2.5D magnetohydrodynamics (MHD) package of the code MPI-AMRVAC, we numerically modelled consecutive CMEs inserted in two different solar winds by imposing shearing motions onto the inner boundary, which in our case represents the low corona. In one of the simulations, the secondary CME was a stealth ejecta resulting from the reconfiguration of the coronal field. The initial magnetic configuration depicts a triple arcade structure shifted southward, and embedded into a bimodal solar wind. We triggered eruptions by imposing shearing motions along the southernmost polarity inversion line, and the computational mesh tracks them via a refinement method that applies to current-carrying structures, and is continuously adapted throughout the simulations. We also compared the signatures of some of our eruptions with those of a multiple coronal mass ejection (MCME) event that occurred in September 2009 using data from spacecraft around Mercury and Earth. Furthermore, we computed and analysed the $Dst$ index for all the simulations performed.}
        {The observed event fits well at 1$\;$AU with two of our simulations, one with  a stealth CME and the other without. This highlights the difficulty of attempting to use in situ observations to distinguish whether or not the second eruption was stealthy, because of the processes the flux ropes undergo during their propagation in the interplanetary space. We simulate the CMEs propagated in two different solar winds, one slow and another faster one. In the first case, plasma blobs arise in the trail of eruptions. The faster solar wind simulations create no plasma blobs in the aftermath of the eruptions, and therefore we interpret them as possible indicators of the initial magnetic configuration, which changes along with the background wind. Interestingly, the $Dst$ computation results in a reduced geoeffectiveness in the case of consecutive CMEs when the flux ropes arrive with a leading positive $B_z$. When the $B_z$ component is reversed, the geoeffectiveness increases, meaning that the magnetic reconnections with the trailing blobs and eruptions strongly affect the impact of the arriving interplanetary CME.}
        {}
        
        
\maketitle

        
    \section{Introduction} \label{sec:intro}
        
Coronal mass ejections (CMEs) are among the most energetic solar phenomena, where magnetized plasma is expelled into interplanetary space, with masses of up to 10$^{16}$ g \citep{schwenn_review, chen_review_cmes}. A commonly associated triggering mechanism consists of shearing motions of photospheric plasma, usually of opposite magnetic polarities. The frozen-in condition forces the overlying coronal loops to twist, become unstable, reconnect, and erupt as CMEs. Most of the time during solar maximum, these ejections are related with other solar events or features such as flares, coronal dimmings, filament eruptions, extreme ultraviolet (EUV) waves, or post-flare loops \citep{hudson_cme_sources}, which are good indicators of the source region of the CMEs. However, during solar minimum the percentage of eruptions correlated with a clear source region is much lower \citep{ma_stealth}, as is the overall number of CMEs. Those events observed in coronagraph images but difficult to trace back to their origin without any image processing techniques are called `stealth' CMEs. Such an eruption was first studied by \citet{robbrecht_stealth}, who described a slow streamer blowout that had no obvious low coronal signatures. The study of \citet{elke_stealth} confirmed a characteristically slow nature for these stealth CMEs, along with a small angular width, by analysing a sample of 40 events. \citet{alzate_sources} applied image-processing techniques to the list identified by \citet{elke_stealth} and found some form of low coronal signatures for most of them, concluding that the lack of characteristic features from this type of eruptions is the result of observational limitations. \citet{erika_stealth} also employed a multitude of imaging and geometric techniques, using various spacecraft images to identify the source of four stealth CMEs occurring throughout all the stages of the solar cycle. Several other authors such as \citet{kilpua_stealth} and \citet{nariaki_stealth} studied comprehensive sets of stealth CMEs, as well as their interplanetary counterparts, and even their effect on Earth's magnetosphere, which reached the level of intense geomagnetic storm.\\ 
A particular subset of these eruptions consists of stealthy blob-like structures that trail a CME whose source region could be identified. This kind of configuration has been numerically simulated by \citet{zuccarello}, \citet{bemporad}, and \citet{talpeanu}, who were motivated by a multiple coronal mass ejection (MCME) observed on 21-22 Sept 2009. In the latter paper (from now on referred to as paper I), the authors debate whether or not the second eruption is indeed a stealth CME, because their simulations are more consistent with a scenario of an eruption driven by shearing motions from the inner boundary.\\
The present study is a follow-up of paper I; we model the propagation of their multiple eruptions to 1 AU in different background solar winds, compare in situ signatures, and compute the hypothetical induced geoeffectiveness. Numerical simulations are important in studying the propagation of multiple consecutive CMEs, because it is known that these can interact with each other and also with the background solar wind via magnetic reconnection and deflection \citep[e.g.][]{manchester_cmes_interaction}. Most of the time, the observational resources available to study these processes consist of remote-sensing images and in situ data at 1 AU. The limited information accessible between Sun and Earth means that numerical simulations are extremely useful in further understanding these interactions, as well as anticipating the morphology of interplanetary CMEs (ICMEs) arriving at Earth. Motivated by the above, we simulate the propagation of consecutive slow CMEs inserted in different solar winds in an attempt to further understand how such eruptions interact, and how the magnetic structure is distorted during the propagation. If these factors are reliably modelled, then it may be possible to compute the geomagnetic impact of the CMEs hours or even days ahead. We will assess this impact using the $Dst$ index, which measures changes in the horizontal component of the magnetic field at ground level \citep{sugiura_dst}. Several authors have developed ways of $Dst$ prediction using solar wind parameters, which take into consideration different effects and mechanisms, from the first and simplest model of \citet{burton_dst_1975} to some of the most complex semi-empirical algorithms of \citet{temerin&li_2002,temerin&li_2006}. In the final step of our analysis, we use a modified version of the method outlined by \citet{dst_computation} to compute the geoeffectiveness of our simulated CMEs. We chose this method because of its relative simplicity and fast computational speed. 
      
        
    \section{Observations} \label{sec:observations} 
        
\begin{figure}[h!]
        \centering
        \includegraphics[width=0.8\columnwidth]{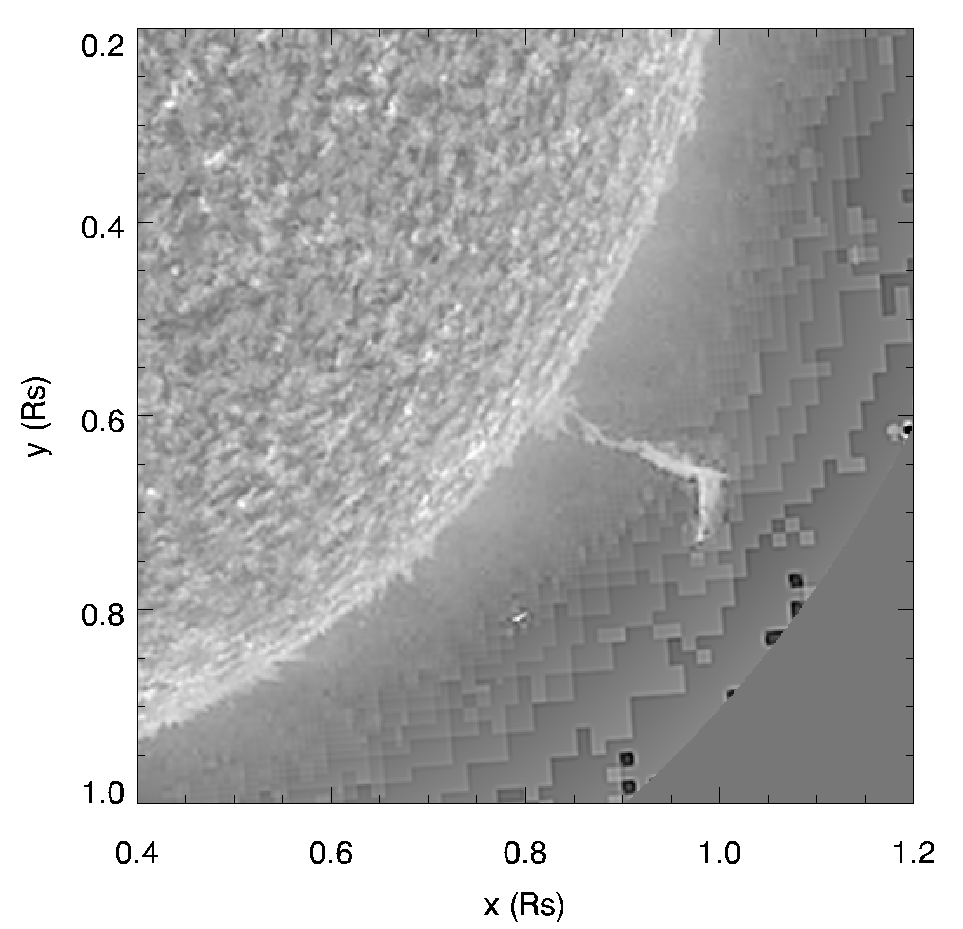}
        \caption{Image from STEREO-B EUVI (304\AA) of the erupting prominence, taken on 21 September 2009 at 16:36UT. The image was scaled to enhance the prominence and a mask was applied from 1.35 R$_{\sun}$ outwards to remove the large noise at the edges. This figure was reproduced with permission from \citet{talpeanu}.}
        \label{fig:prominence}
\end{figure}

The observed event modelled here consists of a multiple coronal mass ejection \citep[MCME;][]{bemporad} that was seen on 21-22 Sept 2009. The source of the first CME (hereafter CME1) was an Earth-directed small prominence eruption seen on the western limb of the Sun by SECCHI-Extreme UV Imager \citep[EUVI;][]{secchi} on board the STEREO-B spacecraft. At that time, the angular separation between STEREO-B and Earth was 55.6$\degr$. The prominence departed from an approximate latitude of 37$\degr$ south and exited the EUVI field of view at 19:37 UT on 21 Sept 2009 (Fig.~\ref{fig:prominence}). As STEREO-B was east of Earth, the right-hand side of Fig. \ref{fig:prominence} is Earth-directed, as is the prominence. During its eruption, this slow CME strongly deflected northward and reached the equatorial plane within several solar radii of the surface of the Sun. The reader can find detailed EUV and coronagraph observations and kinematics information of this event in paper I. Underlying the prominence was an active region at approximately 38\degr S and 15\degr W (as seen from Earth), although it did not have a NOAA number assigned. We performed the visual analysis from the vantage point of STEREO-B, as the event showed no clear signatures as seen from Earth. Almost 8 hours after CME1, on 22 Sept 2009 at $\sim$04:05 UT, a second eruption (hereafter CME2) appeared in the COR1-B \citep{thompson-cor, stereo_secchi} coronagraph field of view at a height of $\approx2R_S$. 
After a thorough investigation of the remote-sensing images in different wavelengths, the authors of paper I could not indicate a clear source region for CME2. This reason, along with the fact that CME2 was only seen at high distances from the solar surface, motivated these latter authors to consider the second eruption a stealth CME. Despite having a smaller angular width and being weaker in white-light brightness as compared to CME1, CME2 followed a similar path to its predecessor and was also strongly deflected towards the equator. This behaviour was explained and numerically simulated by \cite{zuccarello} and in paper I, where a more detailed description of the initiation phase of the eruptions can be found.\par
\begin{figure}[h!]
                \centering
                \includegraphics[width=1\columnwidth]{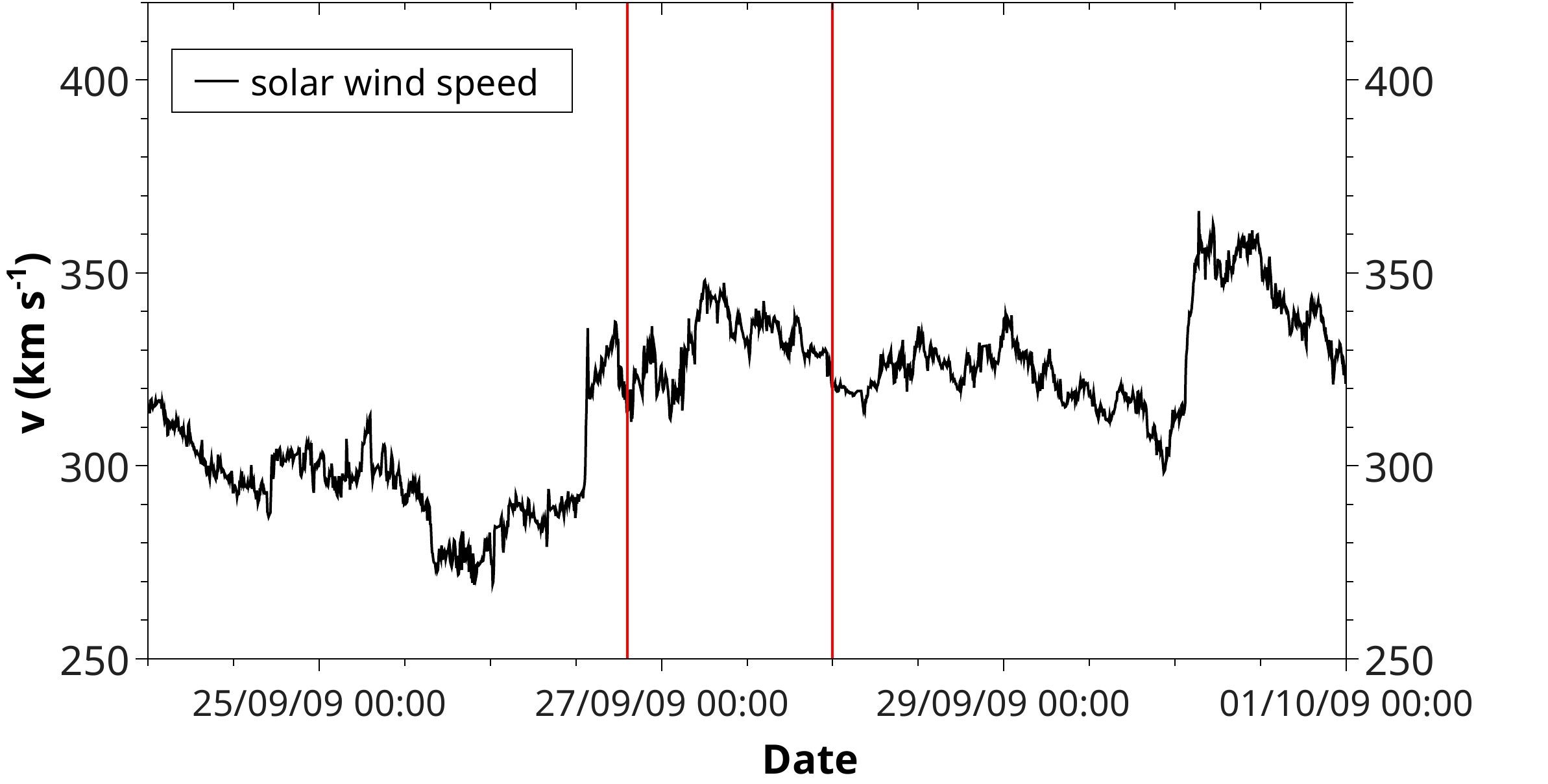}
                \caption{Solar wind speed recorded at Earth, taken from the OMNI database (https://omniweb.gsfc.nasa.gov/). The two red lines indicate the expected arrival time interval of the two CMEs.}
                \label{fig:speed_5min_omni}
                \vspace{-0.3cm}
        \end{figure}

 
In paper I, the deprojected velocities of the CMEs have been calculated along with the propagation longitudes, and the results were 257 $\pm$ 69 km s$^{-1}$ and 5.82\degr W for CME1, and 349 $\pm$ 70 km s$^{-1}$ and 6.7\degr W for CME2. The second eruption was faster than the first one because of the depletion of solar wind material caused by the passage of its precursor. As both ejections were Earth-directed and there were no major CMEs before or after them, one can assume that in the absence of strong erosion forces during their propagation, the flux ropes could have arrived at our planet. Their low speeds also led to the expectation that the CMEs would arrive at Earth with almost the same speed as the solar wind into which they were inserted because of the drag forces exerted onto them. This speed was calculated in paper I, and was assumed to remain constant throughout the propagation, resulting in an average of $\approx$330 km s$^{-1}$ (Fig.~\ref{fig:speed_5min_omni}) and providing an arrival date at 1 AU of between 27 and 28 Sept 2009. In the present follow-up paper, we investigate possible ICME signatures at Mercury and Earth, and find a clear jump in total magnetic field, as well as a smooth rotation of the B$_z$ component in MESSENGER data. As it was such a weak event, the signatures at 1 AU were not as clear, but still distinguishable from the background noise, and they are presented and compared with our simulations in Section \ref{sec:insitu}.  

    \section{Numerical MHD code and methods} \label{sec:simulations}

In this follow-up study, our goal is to propagate the CMEs simulated in paper I  out to 1 AU, and therefore the numerical setup and code are very similar. In addition, we investigate the effect of an increase in speed of the background solar wind on the dynamics and resulting geoeffectiveness of the eruptions, which we discuss further below. We performed numerical magnetohydrodynamics (MHD) simulations using the Message Passing Interface - Adaptive Mesh Refinement - Versatile Advection Code \citep[MPI-AMRVAC;][]{keppens_amrvac, porth_amrvac, xia_amrvac}, with a 2.5D spherical axisymmetric solution. The computational domain spans from the low corona until 1.5AU, and from the north solar pole to the south solar pole, that is (\emph{r}, $\theta$) $\in$ [1, 322] R$_{\sun}$ $\times$ [0\degr, 180\degr], where $r$ is the radial distance from the centre of the Sun and $\theta$ is the heliographic colatitude. We extend the outer boundary by 0.5AU as compared to the previous study because here we analyse the in situ signatures at Earth and we want to avoid possible artificial boundary effects. This change does not influence the initiation phase of the CMEs or the early propagation phase discussed in paper I. The 2D logarithmically stretched grid has an initial resolution of 516 $\times$ 240 cells in the $r$ and $\theta$ directions, and the number of cells is increased by up to two fold through an adaptive mesh refinement (AMR) protocol that tracks the electrical-current-carrying structures which potentially appear at magnetic reconnection sites. This refining method is performed via a parameter used by \citet{karpen} and \citet{skralan} that is defined as follows: 

\begin{equation}
        c \equiv \frac{|\iint_S \nabla \times \boldsymbol{B} \cdot \mathrm{d} \boldsymbol{a}|}{\oint_C |\boldsymbol{B} \cdot \mathrm{d} \boldsymbol{l}|} = \frac{|\oint_C \boldsymbol{B} \cdot \mathrm{d} \boldsymbol{l}|}{\oint_C |\boldsymbol{B} \cdot \mathrm{d} \boldsymbol{l}|} = \frac{|\sum_{n=1}^{4} B_{t,n} l_n|}{\sum_{n=1}^{4} |B_{t,n} l_n|}, \label{c_param}
\end{equation} 

\noindent where $B_{t,n}$ is the tangential component of the magnetic field along the segment $l_n$, which is the length of each side of a grid cell. The term on the right-hand side represents the discrete form of the middle term and each member of the fraction is the sum of the products between $B_{t,n}$ and the length of the edges $l_n$ along all four sides of a cell. The line integral at the numerator of the middle fraction is obtained by applying Stokes' theorem to the surface integral in the left-hand side term, and this is calculated along the curve (contour) of a grid cell. The surface integral is calculated on the surface of a grid cell. As the only difference between numerators and denominators is the place where the absolute value is applied, this results in $c$ being the ratio of the magnitude of the electrical current passing through the surface $S$ of contour length $C$ to the sum of the absolute value of all of its components. Therefore, the parameter $c$ can have values of between 0 and 1 depending on the magnetic field and its non-potentiality. Depending on the magnitude of $c$ in each block, the grid is either refined for values of $c$ > 0.02, or coarsened to a lower level if $c$ < 0.01, because there are no strong current-carrying structures in that region and there is no need to use extra computational resources. If $c$ lies in the interval delimited by these two values, then the grid remains at the resolution imposed by the AMR routine in the previous time-step. The blocks are also refined to the maximum of two levels close to the inner boundary in a region that encompasses the coronal magnetic structures in order to avoid diffusivity changes and artificial dynamics during the initiation phase of the CMEs. This region is dependent on the background solar wind and is defined regardless of the parameter $c$. The logarithmic stretching of the grid keeps the scale of the cells constant, and the ratio between the widths or heights of the furthest cell to those of the closest cell in the same grid level of refinement is $\approx$321. \par           
The MHD equations are spatially discretized using the total variation diminishing Lax-Friedrichs (TVDLF) scheme, and temporally discretized using a two-step predictor corrector suitable for the aforementioned finite-volume method. The slope limiter minmod was used in combination with a Courant-Friedrichs-Lewy (CFL) number of 0.3, giving a very diffusive and stable configuration. In order to maintain a divergence-free magnetic field solution, a new variable is introduced in the system that smooths and transports the unphysical monopoles that might be created to the outer boundary of the computational domain; this method is called the generalised Lagrange multiplier \citep[GLM; ][]{glm}.\par
The initial conditions into which we are erupting CMEs consist of a bimodal background solar wind symmetric in the $\phi$ direction and obtained by introducing extra source terms to the momentum and energy equation that account for gravity and heating mechanisms. The bimodality is expressed as a function of latitude, with faster solar wind at high latitudes, consistent with the data obtained by the Solar Wind Observations Over the Poles of the Sun \citep[SWOOPS,][]{swoops_ulysses} instrument on board Ulysses spacecraft \citep{ulysses_solar_wind}, as well as with interplanetary scintillation \citep[IPS,][]{interplanetary_scintillation} observations at solar minimum. This type of solar wind model was used by \citet{jacobs_2005}, \citet{chane_2006}, \citet{chane_2008}, and \citet{skralan_2019}, and in paper I. The volumetric heating function that defines the separation between slow and fast wind was introduced by \citet{groth} and \citet{manchester} and has the following empirical form:   
        
\begin{equation}
        Q = \rho q_0 (T_0 - T) \exp \left[ -\frac{(r-1\mathrm{R_{\sun}})^2}{\sigma_0^2} \right],  
\end{equation}

\noindent where $\rho$ is the mass density, $q_0$ is the amplitude of the volumetric heating and has a value of $10^{6}$ ergs g$^{-1}$ s$^{-1}$ \rm{K}$^{-1}$, $T\,\mathrm{(K)}$ is the temperature, and $r$ (R$_{\sun}$) is the distance from the centre of the Sun. The parameter $\sigma_0$ (R$_{\sun}$) represents the heating scale height, and is defined as a function of the value of a critical angle, $\theta_0$ (measured from the North pole), as follows: $\sigma_0=4.5[2-\sin^2(\theta)/\sin^2(\theta_0)] \,\mathrm{R_\sun}$. For a more accurate solar wind description, $\theta_0$ is also dependent on the distance from the Sun, as follows:

\begin{dmath}
\mathrm{sin}^2(\theta_0)= 
\begin{cases}
    \sin^2(17.5 \degr)+\cos^2(17.5 \degr) (r-1\mathrm{R_{\sun}})/8\mathrm{R_{\sun}},\text{ for } r < 7 \mathrm{R_{\sun}}\\
    \sin^2(61.5 \degr)+\cos^2(61.5 \degr) (r-7\mathrm{R_{\sun}})/40\mathrm{R_{\sun}}, \text{for } 7 \mathrm{R_{\sun}} \leq r < 47 \mathrm{R_{\sun}}\\
    1,              \text{ for } 47 \mathrm{R_{\sun}} \leq r.
\end{cases}
\end{dmath}

\begin{figure}[h!]
        \centering
        \includegraphics[width=0.9\columnwidth]{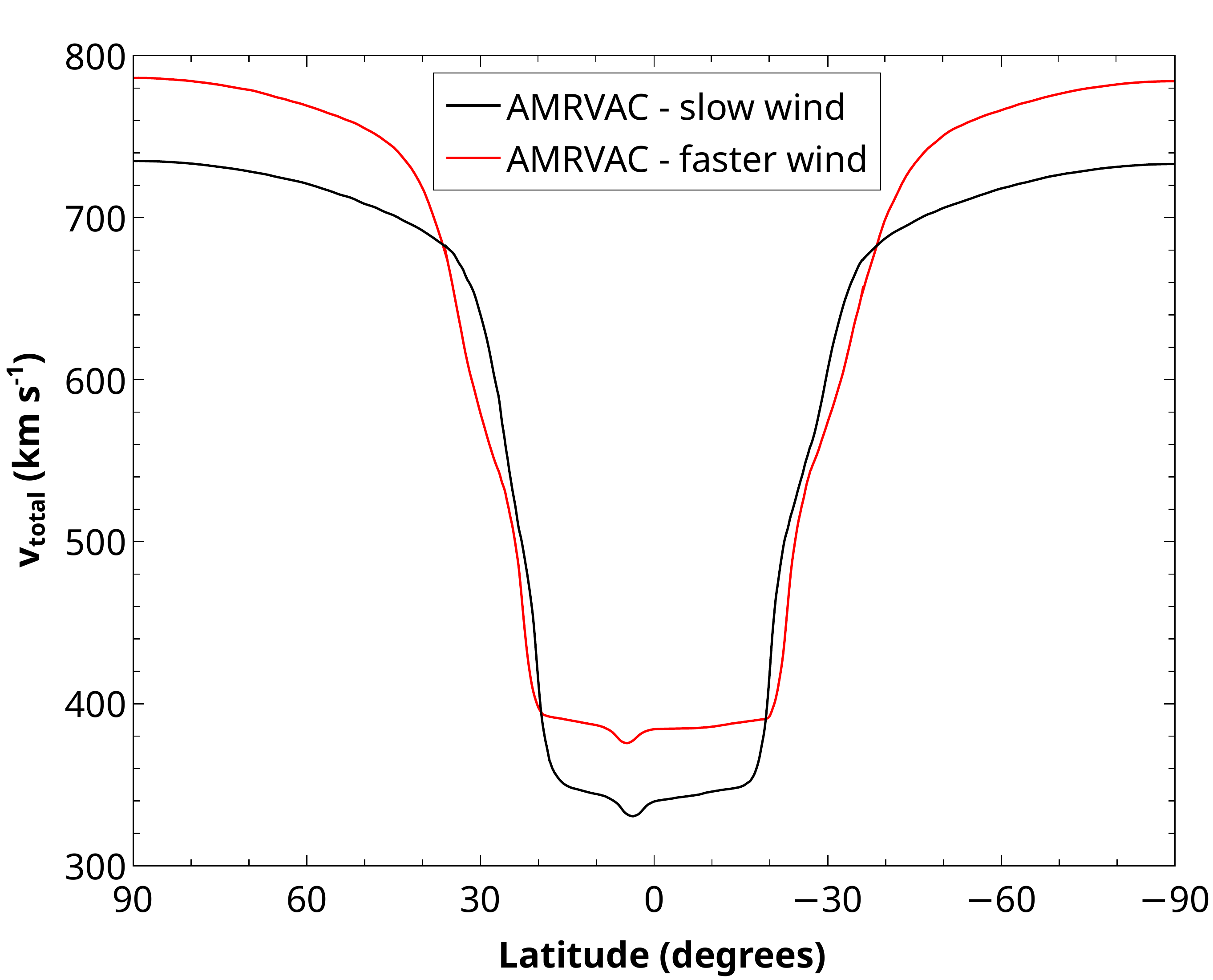}
        \caption{Total speed of the simulated slow (black line) and faster (red line) background solar winds, calculated at 1 AU.}
        \label{fig:speeds_1AU}
\end{figure}

The parameter $T_0 \,\mathrm{(K)}$ represents the target temperature through which we can adjust the momentum of the background wind, and therefore its speed as well. We used the same temperature as in paper I in order to propagate the same solar eruptions, but also a higher value in order to obtain a separate faster solar wind. The aim here is to investigate whether the plasma blobs occurring in the aftermath of eruptions are influenced by the speed of the background wind, but also whether or not the initial magnetic configuration, eruption dynamics, propagation, and geoeffectiveness of CMEs are affected in any way. Therefore, we refer from now on to our two configurations and separate simulations as slow wind and faster wind, and not as composite latitudinal parts of the same solar wind. The two individual types of winds (separate simulations) are determined by $T_0$ in the following way: from $\theta_0$ towards the equator, $T_0=1.32 \times 10^{6}\,\mathrm{K}$ for the slow wind and $T_0=1.5 \times 10^{6}\,\mathrm{K}$ for the faster wind, and from $\theta_0$ towards the pole, $T_0=2.31 \times 10^{6}\,\mathrm{K}$ for the slow wind, and $T_0=2.625 \times 10^{6}\,\mathrm{K}$ for the faster wind. This results in the following minimum and maximum speeds at 1 AU: 330.6 km s$^{-1}$ and 735 km s$^{-1}$ for the slow wind, and 375.7 km s$^{-1}$ and 786.3 km s$^{-1}$ for the faster wind. The minimum value of the speed is found in the equatorial current sheet, which is shifted northward because of the initial asymmetric magnetic configuration described later on. The maximum value is found at the north pole (90$\degr$ latitude), and the speed profile of both background winds at 1 AU can be seen in Fig.~\ref{fig:speeds_1AU}. 

The mass density and temperature are fixed at the inner boundary to $1.66 \times 10^{-16}\,\mathrm{g \, cm^{-3}}$ for both winds, and to $1.32 \times 10^{6}\,\mathrm{K}$ for the slow wind, and to $1.5 \times 10^{6}\,\mathrm{K}$ for the faster wind. The latitudinal component of the speed ($v_\theta$) is set to zero at the inner boundary, whereas the radial component of the momentum is extrapolated in the ghost cells. The differential rotation of the Sun is also reproduced through the azimuthal component of the speed $v_\phi$. The simple solar dipole field is created by fixing $r^2B_r$ at the inner boundary, and by extrapolating $r^5B_\theta$ and $B_\phi$ from the first inner cell. The variables $r^2\rho$, $r^2\rho v_r$, $\rho v_\theta$, $r v_\phi$, $r^2B_r$, $B_\theta$, $r B_\phi$, and $T$ are also continuous at the outer supersonic boundary. \par 
A very common magnetic field structure present on the Sun resembles a triple arcade system embedded in a helmet streamer; this was previously simulated and used by \citet{karpen} to study a breakout event, and by \citet{bemporad} and \citet{zuccarello} to analyse the dynamics and deflection of the same multiple eruption event as ours. These latter authors modelled this structure through the following vector potential: 

\begin{equation}
        A_\phi = \frac{A_0}{r^4 \sin\theta} \cos^2 \left[ \frac{180\degr (\lambda+11.5\degr)}{2\Delta a}\right], 
        \label{eq:A_phi} 
\end{equation}       

\noindent where $\Delta a = 37.2 \degr$ represents half the width of the arcade system, $\lambda = 90\degr-\theta$ the solar latitude, $A_0 = 0.73\,\mathrm{G \cdot R_{\sun}^5}$, and the entire arcade system is shifted to the south by $11.5\degr$. These values are similar to those obtained by \citet{zuccarello} and were adjusted such that the resultant magnetic configuration resembles the coronal extrapolation they performed from magnetogram data taken on 19 Sept 2009.\\
In the current study, we introduce such an arcade model through the magnetic field components obtained by taking the curl of the vector potential $\mathbf{A}$ defined by Eq.~\ref{eq:A_phi} (the other components of $\mathbf{A}$ are zero):

\begin{eqnarray}
        B_r &=& \frac{A_0}{r^5 \sin\theta} \frac{180\degr}{\Delta a} \cos \left[ \frac{180\degr (\lambda+11.5\degr)}{2\Delta a} \right] \nonumber\\
        &&\cdot \sin \left[ \frac{180\degr (\lambda+11.5\degr)}{2\Delta a} \right],\\
        B_\theta &=& \frac{3A_0}{r^5 \sin\theta} \cos^2 \left[ \frac{180\degr (\lambda+11.5\degr)}{2\Delta a}\right].
\end{eqnarray}       
 
\begin{figure}[h!]
        \centering
                \begin{overpic}[width=1\linewidth]{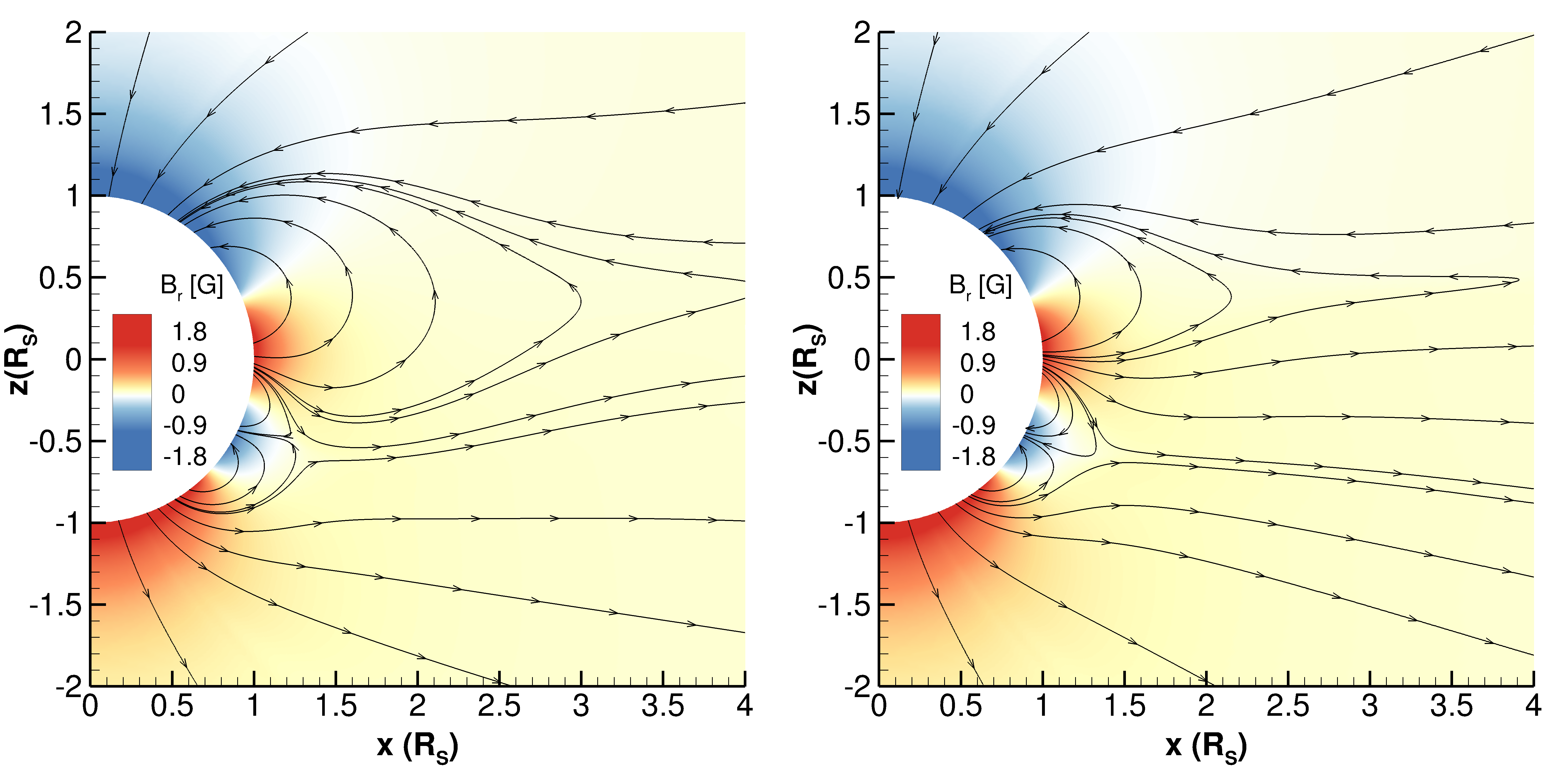}
            \put(30,28.5){\thicklines \oval(5,19)[r]}
            \put(78,31){\color{red} \thicklines \oval(4,9)[r]}            
            \put(72,20){\color{blue} \thicklines \oval(3.5,7)[r]}            
        \end{overpic}
        \caption{Initial magnetic field configuration - $B_r$ (colour scale) and selected magnetic field lines in the meridional plane for the slow solar wind (left side) and faster solar wind (right side).}
        \label{fig:arcades}     
\end{figure}

 These extra components were only added to the dipole magnetic field in the latitude interval $ \lambda \in [-48.7\degr, 25.8\degr]$. This configuration is applied on both solar winds and provides a radial magnetic field strength of 1.8 G (or ${1.8 \times 10^5}$ nT) at the poles, and a maximum arcade strength of 1.57 G (or ${1.57 \times 10^5}$ nT). These values were measured at the first cell of the domain, and even though they are extremely similar for both slow and faster background solar winds, the resultant magnetic configurations are not identical, as can be seen in Fig.~\ref{fig:arcades}. The similarity can be seen clearly in the top left panel of Fig. \ref{fig:Bcomponents_winds}. This interesting result is discussed further below. \par 

In order to propagate two of the eruptions obtained in paper I, we created CMEs in the same manner, by applying the same shearing motions and amplitudes onto the inner boundary in the azimuthal direction, summed with the differential rotation previously mentioned. The additional time-dependent $\phi$ component of the speed was introduced only after the solar winds reached a steady-state solution, and has the following profile:

\begin{equation}
        v_\phi = v_0(\alpha^2-\Delta \theta^2)^2 \sin\alpha \sin[180\degr(t-t_0)/\Delta t],  
\end{equation}

\noindent with $\alpha=\lambda - \lambda_0$. This flow spans over $2 \Delta \theta=17.2\degr$ and is centred at $\lambda_0=-40.7\degr$, which is approximately equal to the latitude of the southernmost polarity inversion line inside the southernmost arcade. The shear starts at $t_0=0$ h and is applied for $\Delta t=16$ h, exhibiting a slow increase and decrease in order to not introduce shocks in the system and reaching a maximum half way through this interval. The scaling factor $v_0$ was chosen such that $v_\phi$ does not exceed $10\%$ of the local Alfv\'{e}n speed in either of the simulations. 


    \section{Simulated eruptions and solar winds} \label{sec:eruptions}    

The slow solar wind case in the current study is almost identical to the background wind in paper I, with the two differences that the outer boundary of the computational domain is extended until 1.5 AU, and that the refined grid area close to the Sun is enlarged such that it encompasses the entire arcade system. However, these changes had no effect on reproducing two of the eruptions in paper I in order to propagate them to Earth. \par
\begin{figure}[h!]
        \centering
        \includegraphics[width=1\columnwidth]{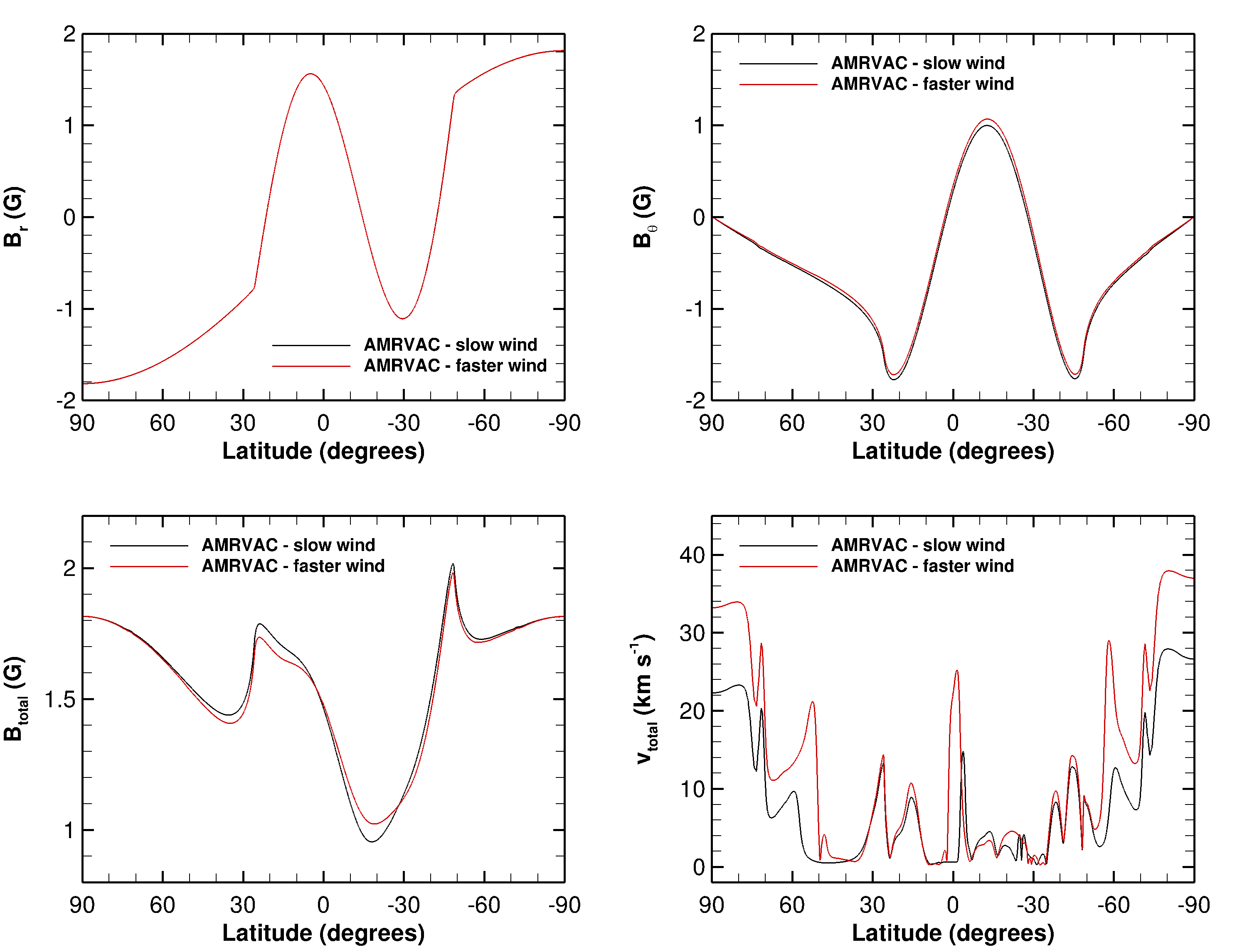}
        \caption{Magnetic field components and total speed calculated in the first cell of the computational domain, for both solar winds: top left - $B_r$, top right - $B_\theta$, bottom left - $B_{total}$, bottom right - $v_{total}$.}
        \label{fig:Bcomponents_winds}   
\end{figure}

 
We obtain an initial interesting result simply from comparing the two background solar wind simulations (Fig. \ref{fig:arcades}). As mentioned above, the same magnetic boundary conditions in combination with a hotter, denser, and faster wind resulted in a very different magnetic configuration, as shown in Fig.~\ref{fig:arcades}. The initial helmet streamer (indicated by the black bracket on the left side of Fig.~\ref{fig:arcades}) breaks up into a northern smaller arcade (red bracket, right-hand side of Fig.~\ref{fig:arcades}) and a southern pseudostreamer (blue bracket, right-hand side of Fig.~\ref{fig:arcades}), as a result of the applied change in temperature and ultimately in speed, as seen in the bottom right panel of Fig.~\ref{fig:Bcomponents_winds}. The values shown in Fig.~\ref{fig:Bcomponents_winds} are extracted from the first cell of the computation domain, and $B_{total}$ and $v_{total}$ are calculated by taking the square root of the sum of the squares of all their respective components. The numerous latitudinal variations in total speed occur because of the plasma changes at the boundaries of each magnetic arcade, and at the two (northern and southern) interfaces between slow and fast wind contained in the same magnetic configuration/simulation. Surprisingly enough, the radial component of the magnetic field is not affected by the modified temperature, as indicated by the overlapping curves in the top left panel of Fig.~\ref{fig:Bcomponents_winds}, and yet a different configuration is created. As a result, the only influence on the total magnetic field, and therefore on the overall arcade configuration, is provided by the $B_\theta$ component (because $B_\phi$ has a value of 0 at the inner boundary), which changes along with other plasma parameters. At higher distances, all the magnetic field components change due to the higher speed and density of the faster solar wind configuration, which determine a different coronal and heliospheric structure.         

Once both solar winds relaxed to a steady state and no more changes occurred in the magnetic field at the outer boundary, we created CMEs by applying an additional $v_\phi$ component to the inner boundary, approximately along the southernmost polarity inversion line as described in the previous section. The shearing motions applied to the arcade in the slow wind case are the same as those that resulted in double and stealth eruptions in paper I, where they are also described in more detail. Briefly, in paper I the authors numerically simulated three different types of consecutive solar eruptions by varying the shearing motions applied at the inner boundary by only 1\%. The two cases of interest for the current study are comprised of a first CME triggered in all the cases by the shearing motions, and a second eruption which was generated either by the shear (double eruption case) or by the reconfiguration of the coronal magnetic field (stealth eruption case). Inside the current sheet that followed the eruptions, plasma blobs were created via magnetic reconnection in all the scenarios. In order to analyse the effects of the trailing eruptions on the first CME, we also performed a simulation with a single flux rope formation by decreasing the amplitude of $v_0$ and implicitly that of $v_\phi$.\par
In numerically simulating a faster solar wind by increasing its temperature, we aim to investigate the occurrence of plasma blobs in the trailing current sheet of CMEs in a different plasma environment. In order to recreate the eruptions that led to the presence of blobs in their aftermath, we applied the amplitudes of $v_0$ that resulted in stealth ejecta and single eruptions in the slow wind case to the inner boundary of the newly simulated faster wind. Below, we refer to these current simulations (same $v_0$ but different background wind) as the stealth speed case (unrelated to the stealth ejecta formation mechanism) and the single eruption case, respectively. Even though the same amplitudes of $v_0$ applied in the ghost cells did not produce the exact same values of $v_\phi$ inside the computational domain, we kept the same numbers of $v_0$ for consistency, and the actual values measured in the first cell of the domain can be seen in Table \ref{table:shearing_speeds}, taken at half (at 8h) the length of shearing time (16 h).

\begin{table}[h!]
        \centering
        \caption[]{Maximum shearing speeds in absolute value for both solar winds in all simulation cases, taken from the first cell of the computational domain.}
        \begin{tabular}{c c c} 
                \label{table:shearing_speeds}
                Background wind             & Eruption type   &  $|v_\phi^{max}| \mathrm{\ [km\ s^{-1}]}$\\
                \hline          
                \noalign{\smallskip}
                \multirow{3}{*}{Slow wind}  & double er.       &  37.43 \\
                                                                    & stealth er.      &  37 \\
                                                                & single er.       &  21.95 \\                                                              
                \noalign{\smallskip}                          
                \hline
                \noalign{\smallskip}
                \multirow{2}{*}{Faster wind} & stealth speed   &  36.77  \\
                                                             & single er.      &  22.33 \\
                \noalign{\smallskip}
                \hline 
        \end{tabular}
\end{table}

\begin{figure*}    
        \centering
        
        \subfloat[$\mathrm{t_{sim}=13 h}$ \label{fig:SW_single_1}]{
                \begin{overpic}[width=0.32\linewidth]{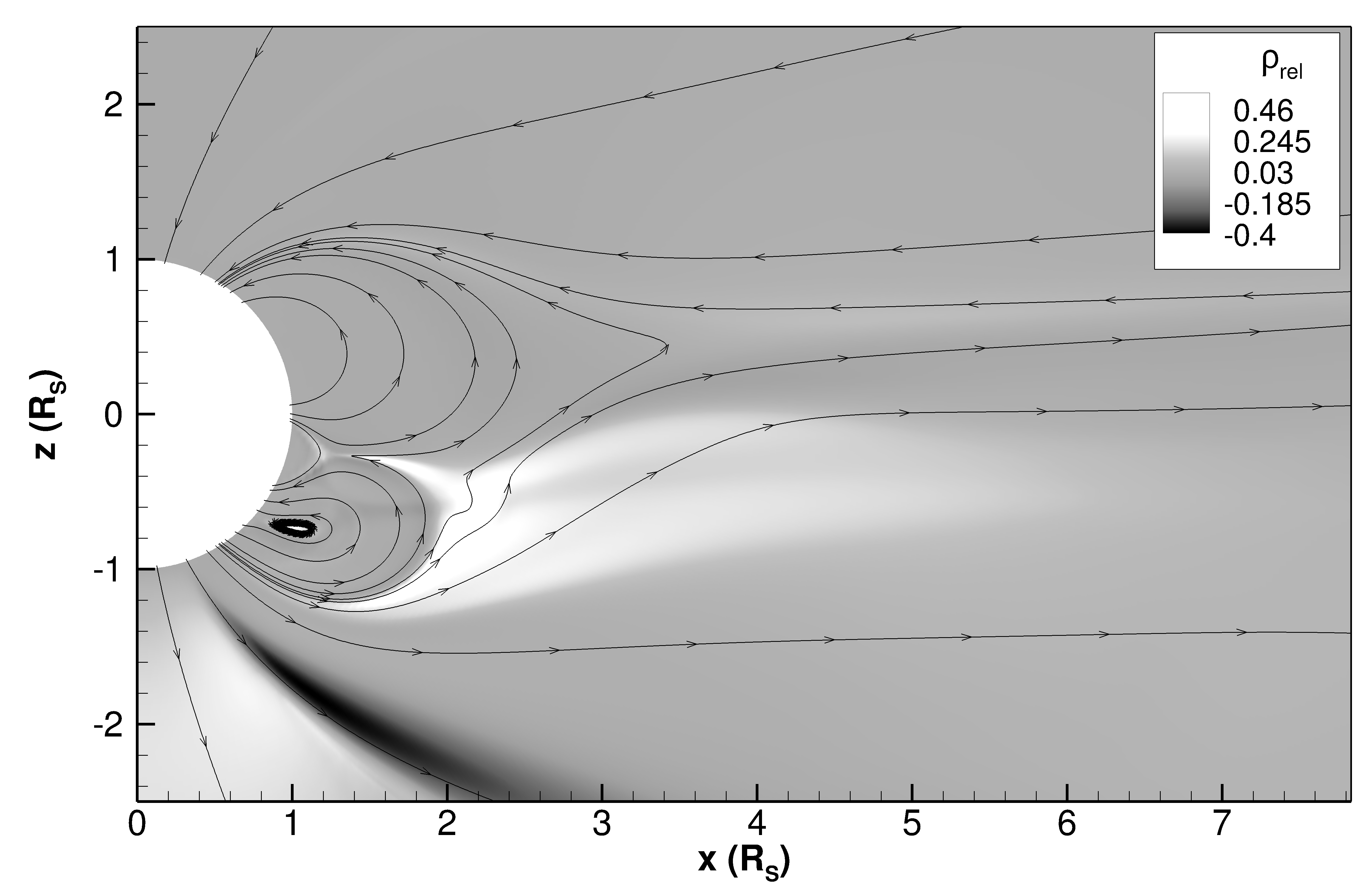}
                        \put(13,58){\tiny \textbf{SW single er.}}
                \end{overpic}}
        \subfloat[$\mathrm{t_{sim}=22 h}$ \label{fig:SW_single_2}]{
                \begin{overpic}[width=0.32\linewidth]{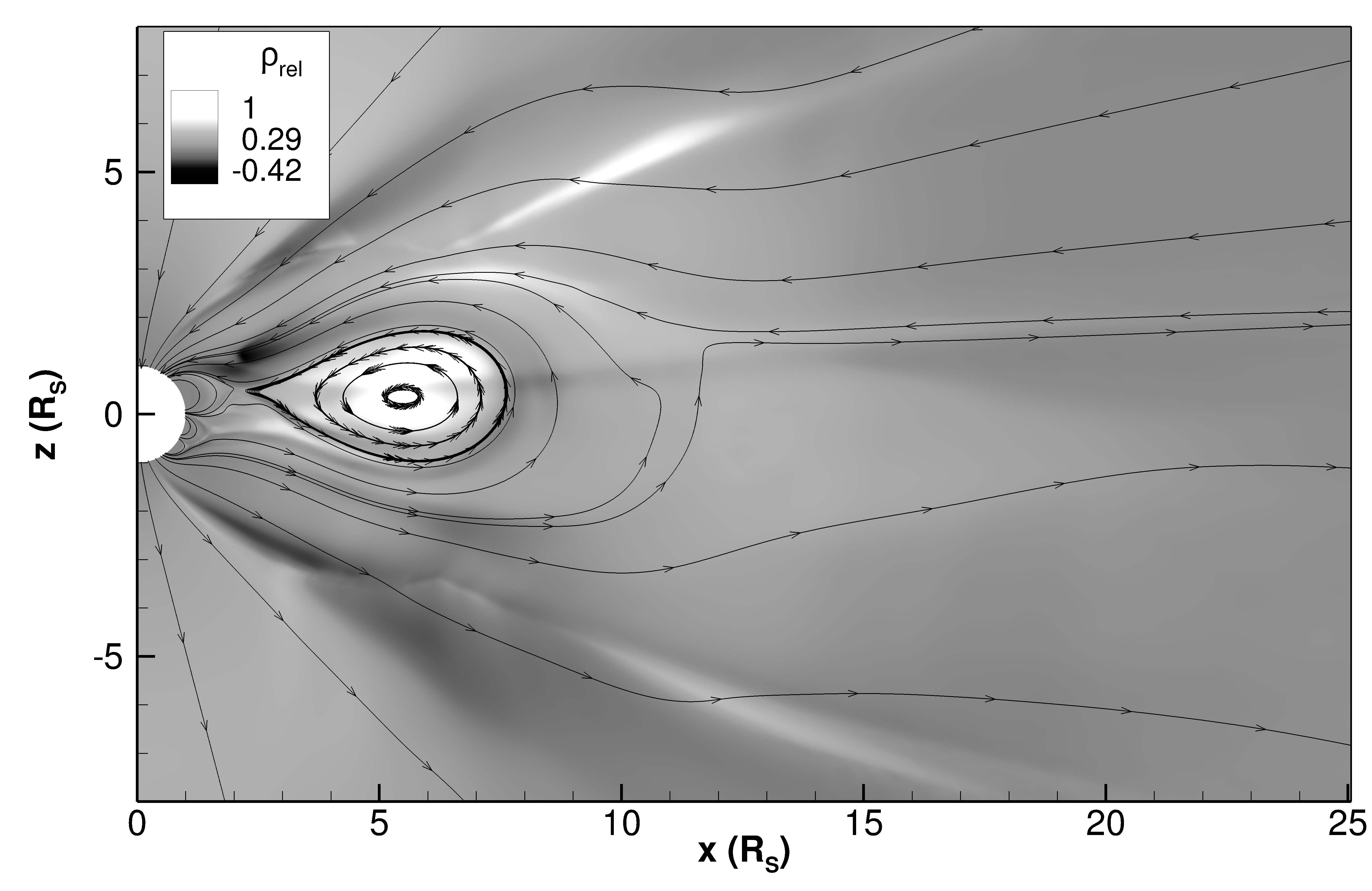}
                        \thicklines
                \end{overpic}}
        \subfloat[$\mathrm{t_{sim}=121 h}$ \label{fig:SW_single_3}]{
                \begin{overpic}[width=0.32\linewidth]{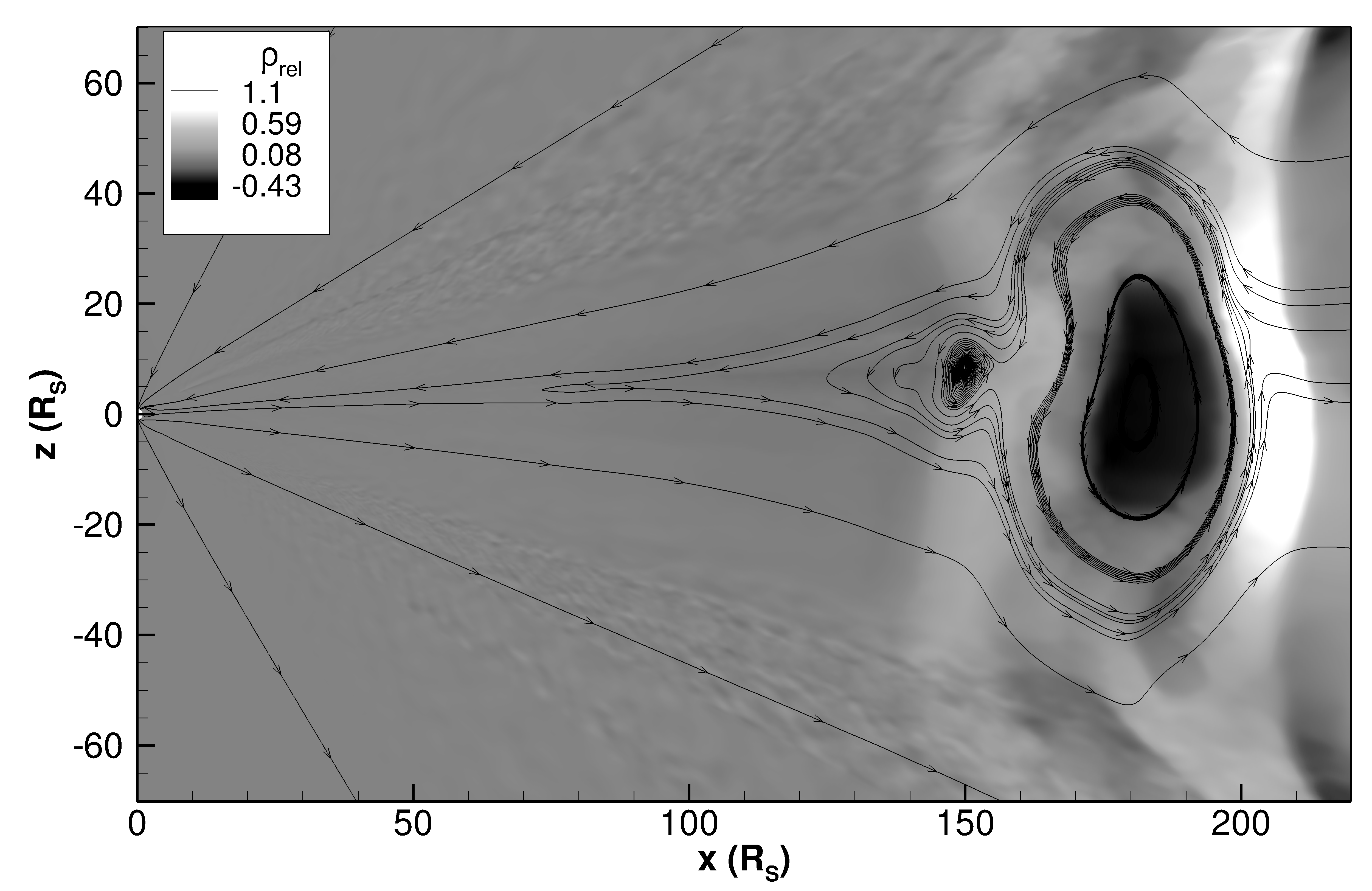}
                        \put(35,28){\vector(1,1){7}}
                        \put(62,49){\color{white} \vector(1,-1){7}}                     
                \end{overpic}}\\
        
        \subfloat[$\mathrm{t_{sim}= 13 h}$ \label{fig:SW_stealth_1}]{
                \begin{overpic}[width=0.32\linewidth]{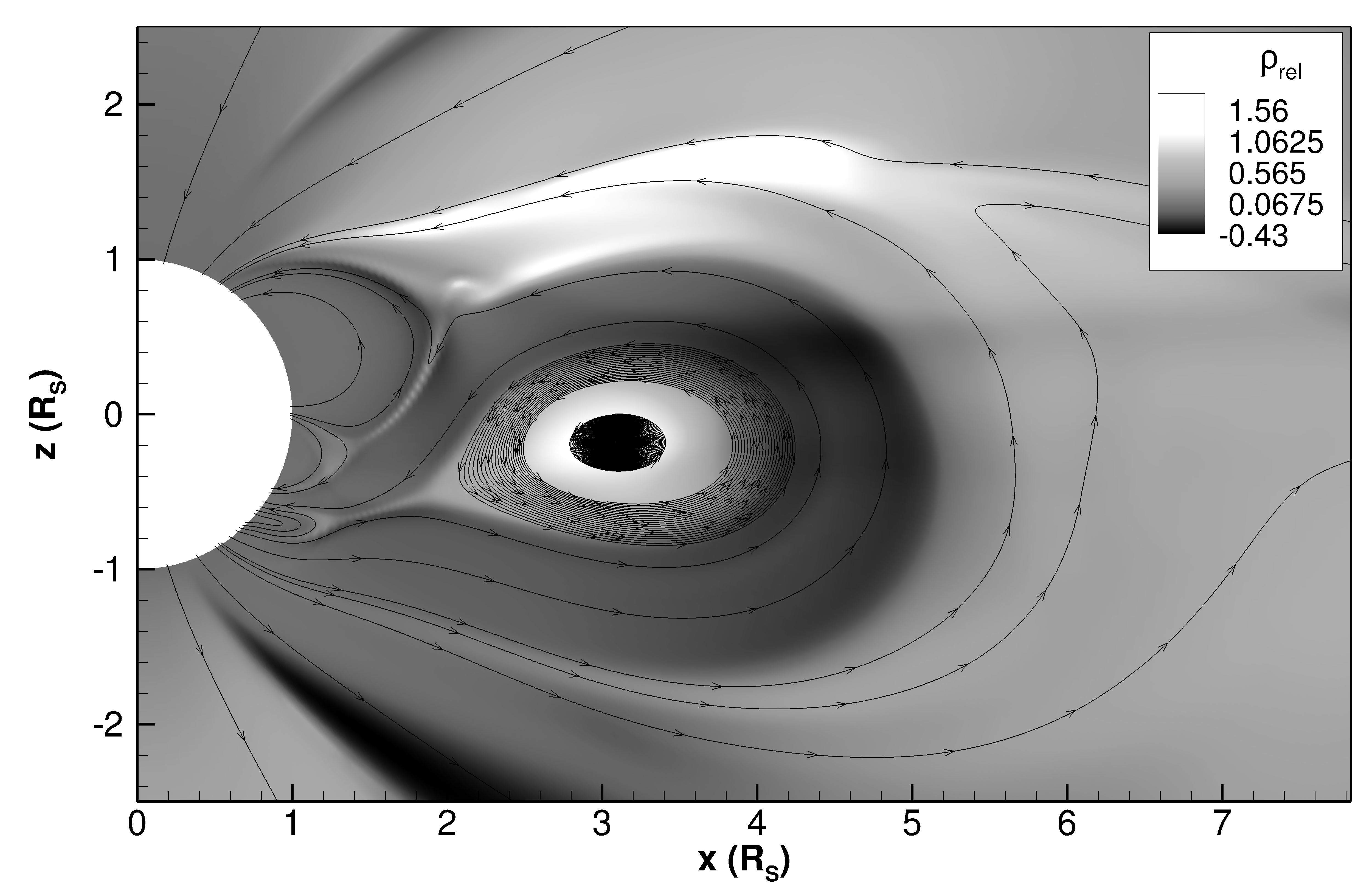}
                        \put(13,58){\tiny \textbf{SW er.+stealth}}                        
                        \thicklines                        
                \end{overpic}}
        \subfloat[$\mathrm{t_{sim}= 22 h}$ \label{fig:SW_stealth_2}]{
                \begin{overpic}[width=0.32\linewidth]{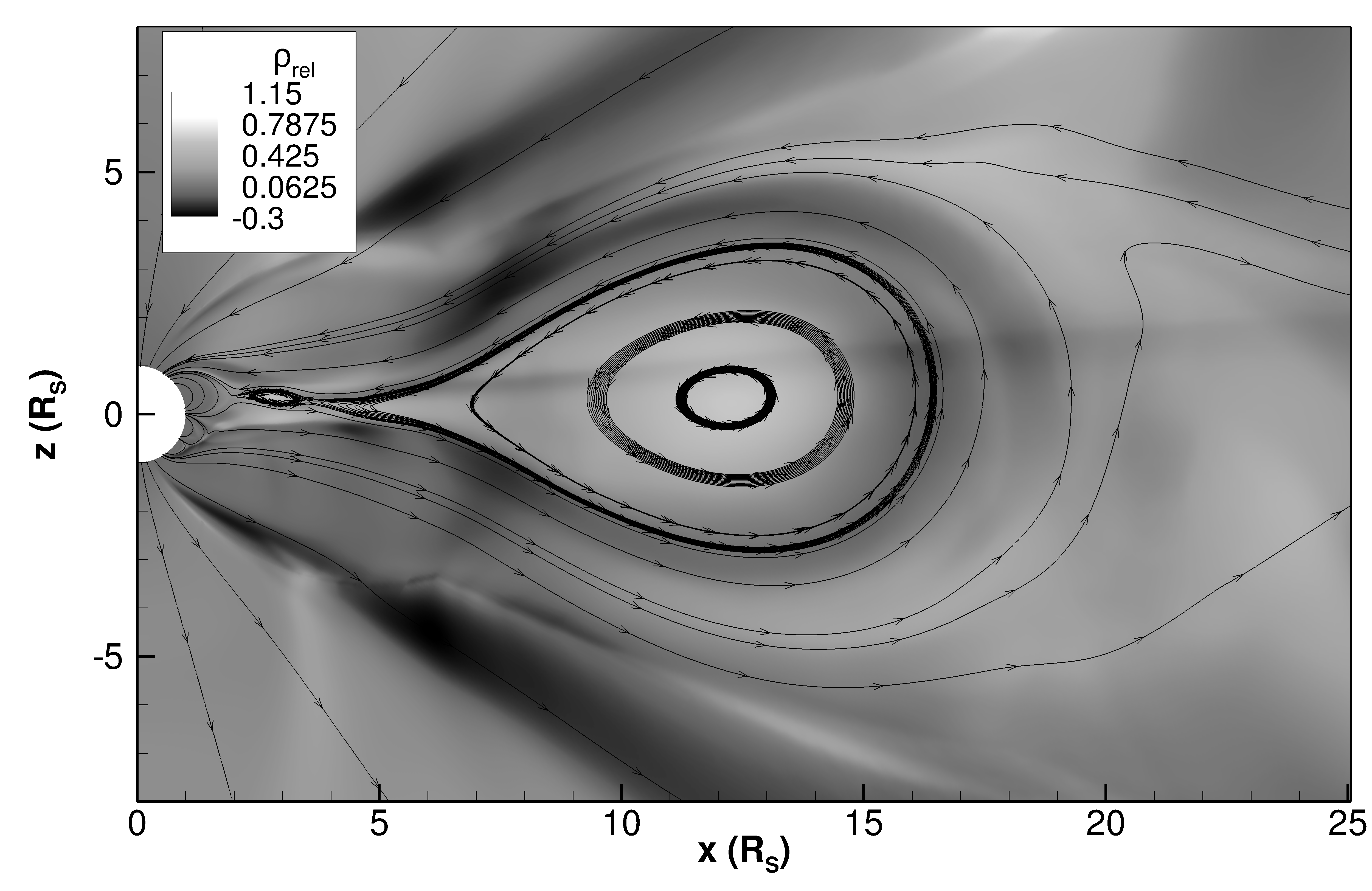}  
                        \put(20,27){\color{white} \vector(0,1){7}}                                              
                \end{overpic}}
        \subfloat[$\mathrm{t_{sim}= 116 h}$ \label{fig:SW_stealth_3}]{
                \begin{overpic}[width=0.32\linewidth]{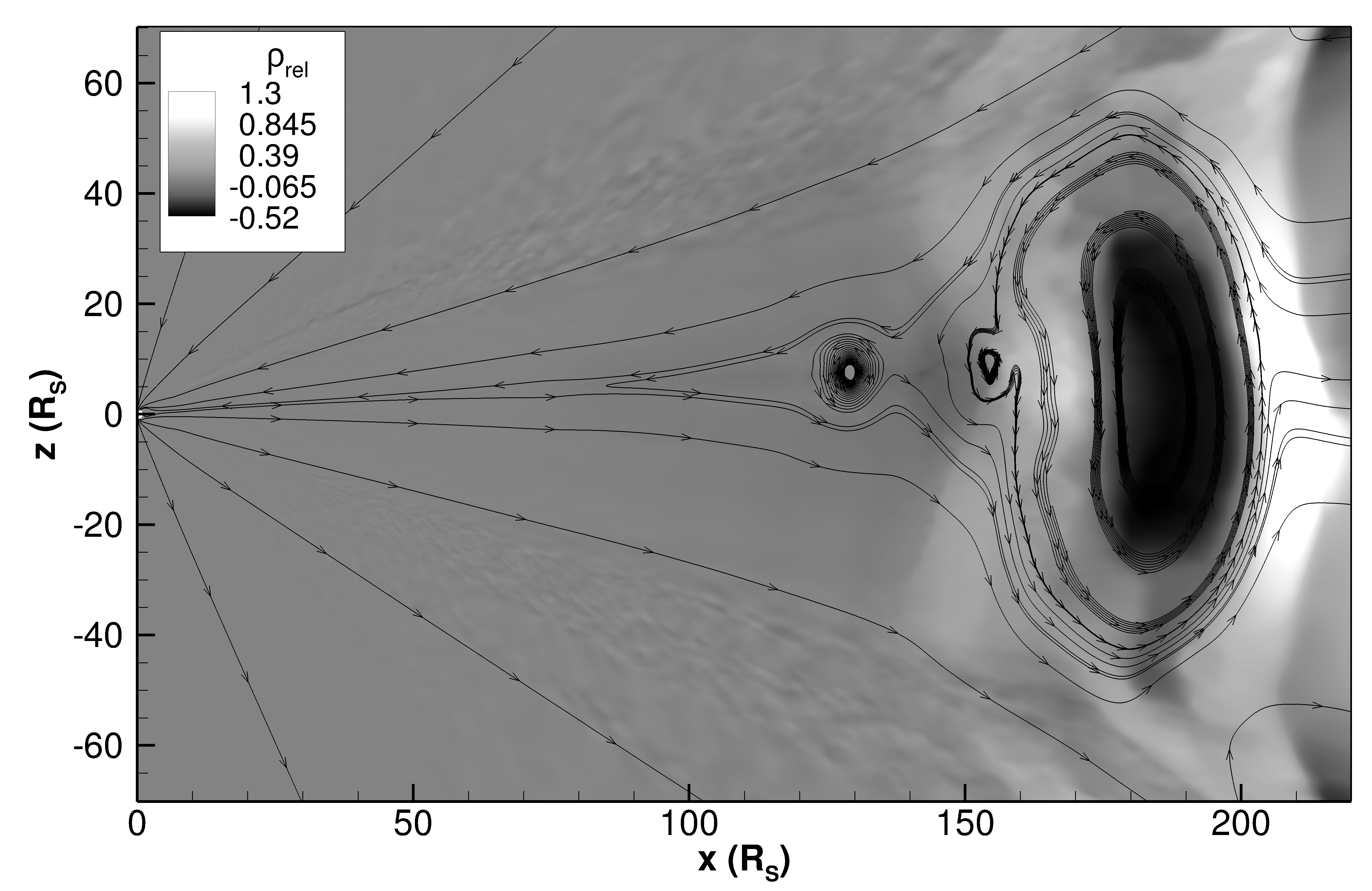}
                \end{overpic}}\\
        \vspace{-0.3cm}         
        
        \subfloat[$\mathrm{t_{sim}= 13 h}$ \label{fig:SW_double_1}]{
                \begin{overpic}[width=0.32\linewidth]{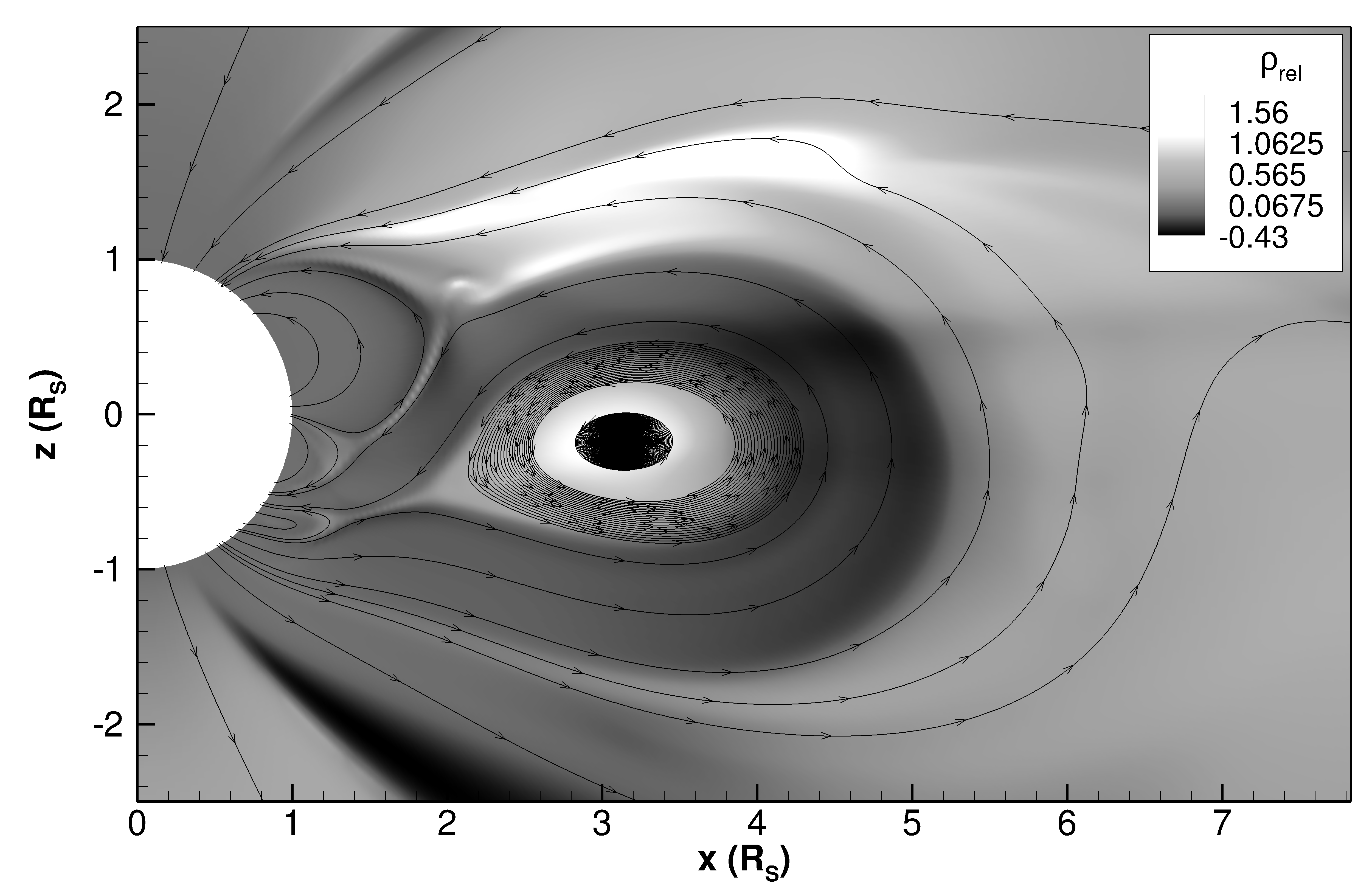}
                        \put(13,58){\tiny \textbf{SW double er.}}                         
                \end{overpic}}
        \subfloat[$\mathrm{t_{sim}= 22 h}$ \label{fig:SW_double_2}]{
                \begin{overpic}[width=0.32\linewidth]{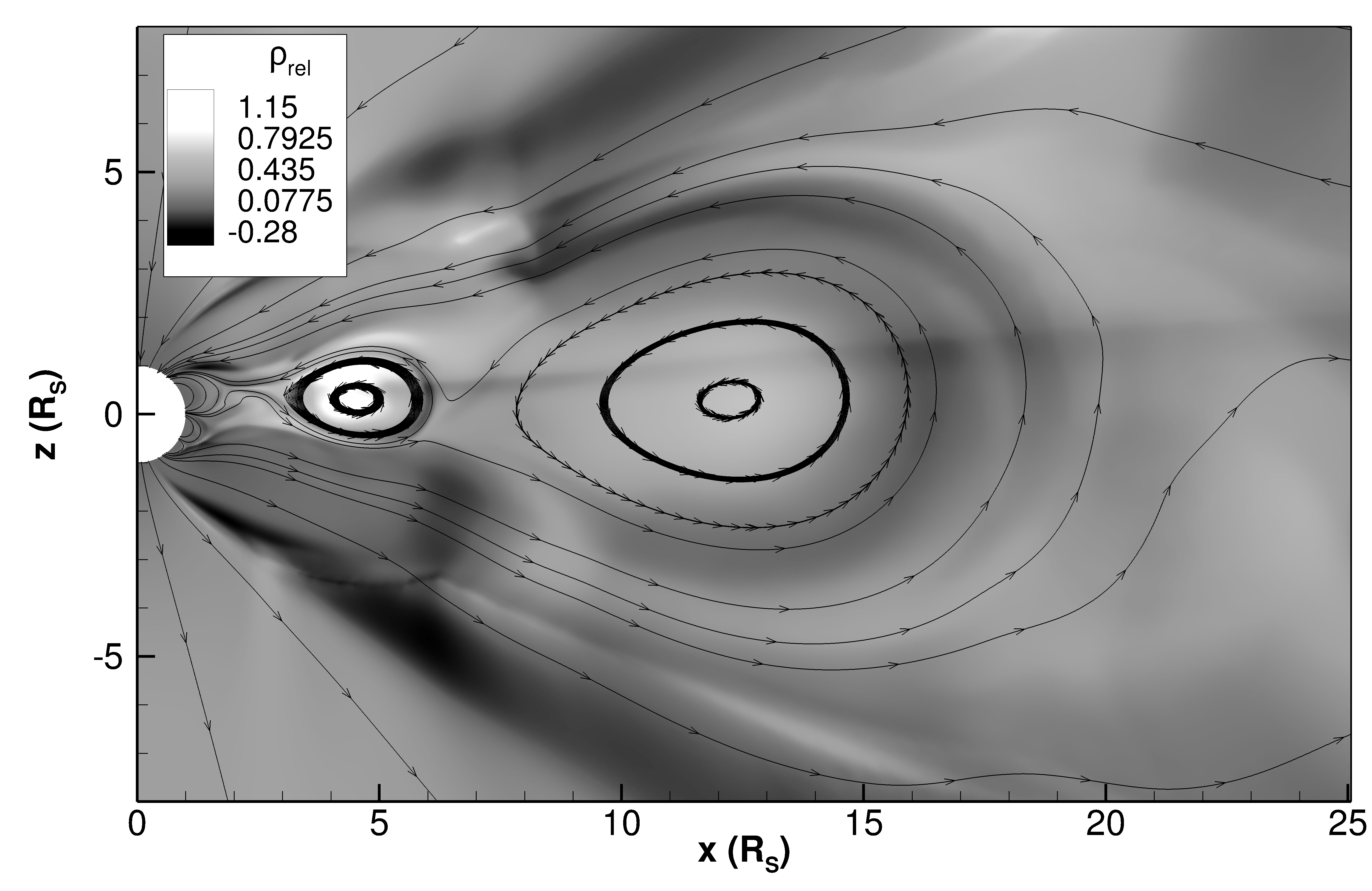}
                \end{overpic}}                 
        \subfloat[$\mathrm{t_{sim}= 115 h}$ \label{fig:SW_double_3}]{
                \begin{overpic}[width=0.32\linewidth]{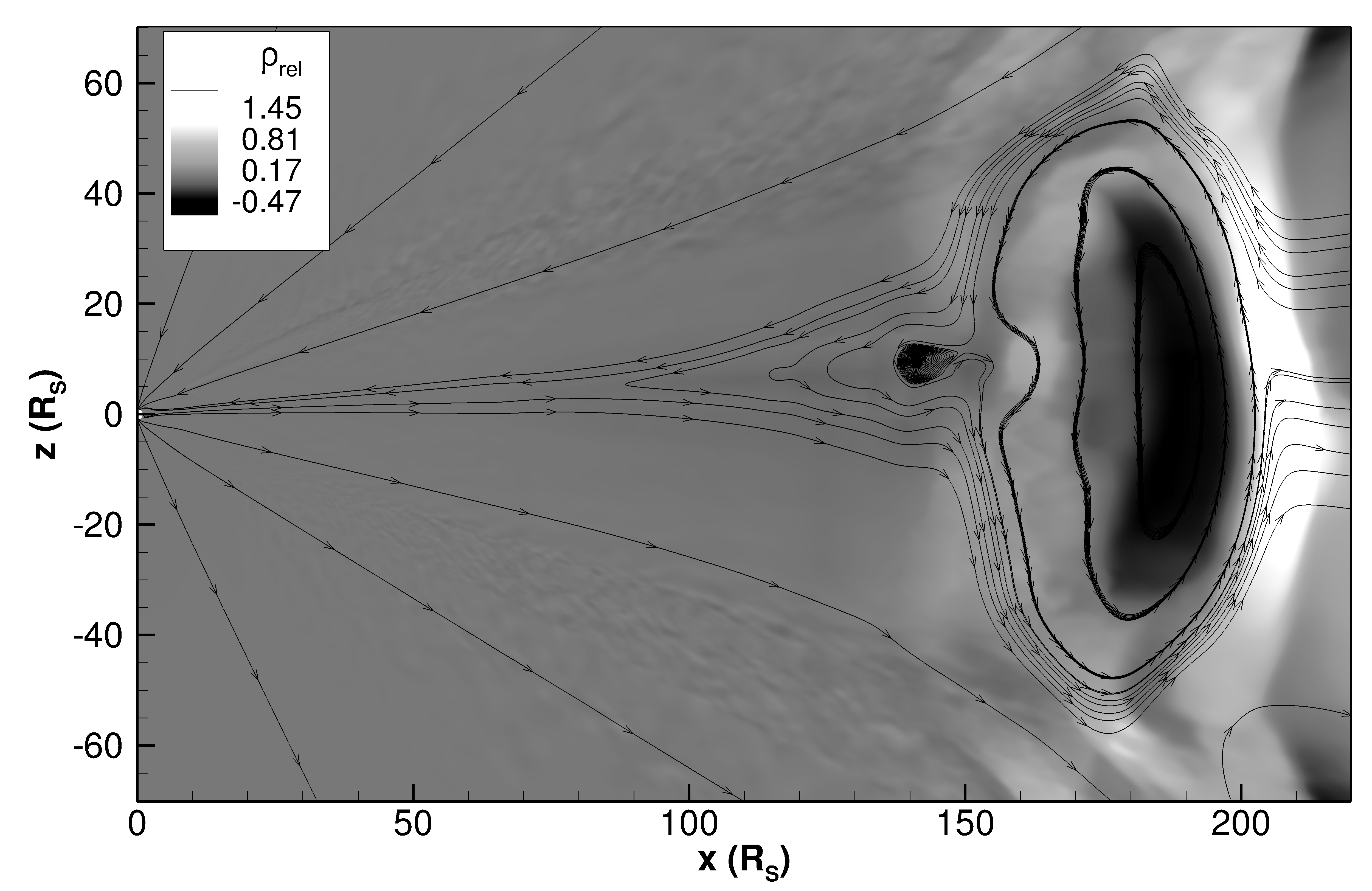}
                \end{overpic}}\\
        \vspace{-0.3cm}
        
        \subfloat[$\mathrm{t_{sim}= 13 h}$ \label{fig:FW_single_1}]{
                \begin{overpic}[width=0.32\linewidth]{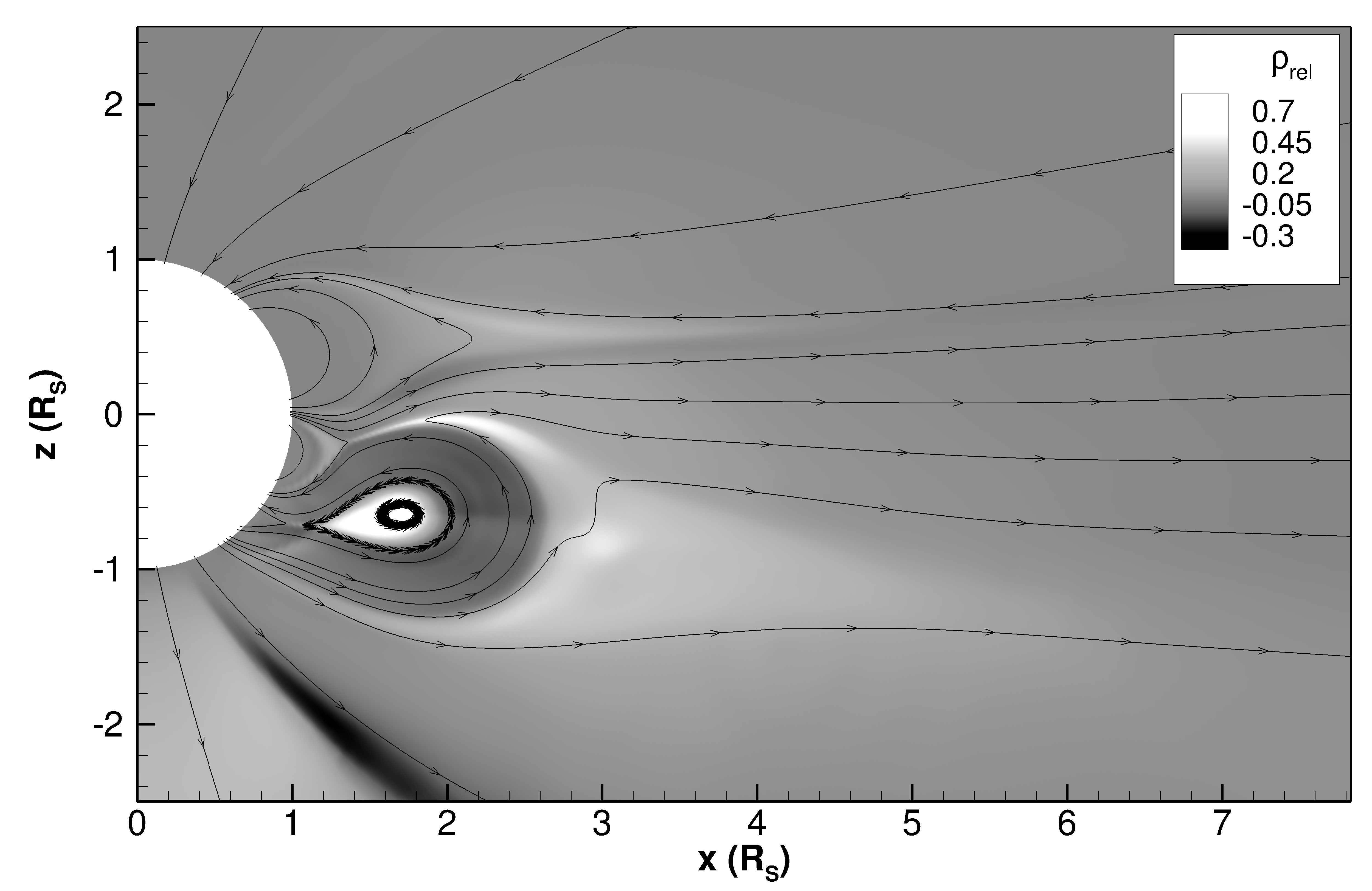}
                        \put(13,58){\tiny \textbf{FW single er.}}
                        \thicklines                          
                \end{overpic}}
        \subfloat[$\mathrm{t_{sim}= 22 h}$ \label{fig:FW_single_2}]{
                \begin{overpic}[width=0.32\linewidth]{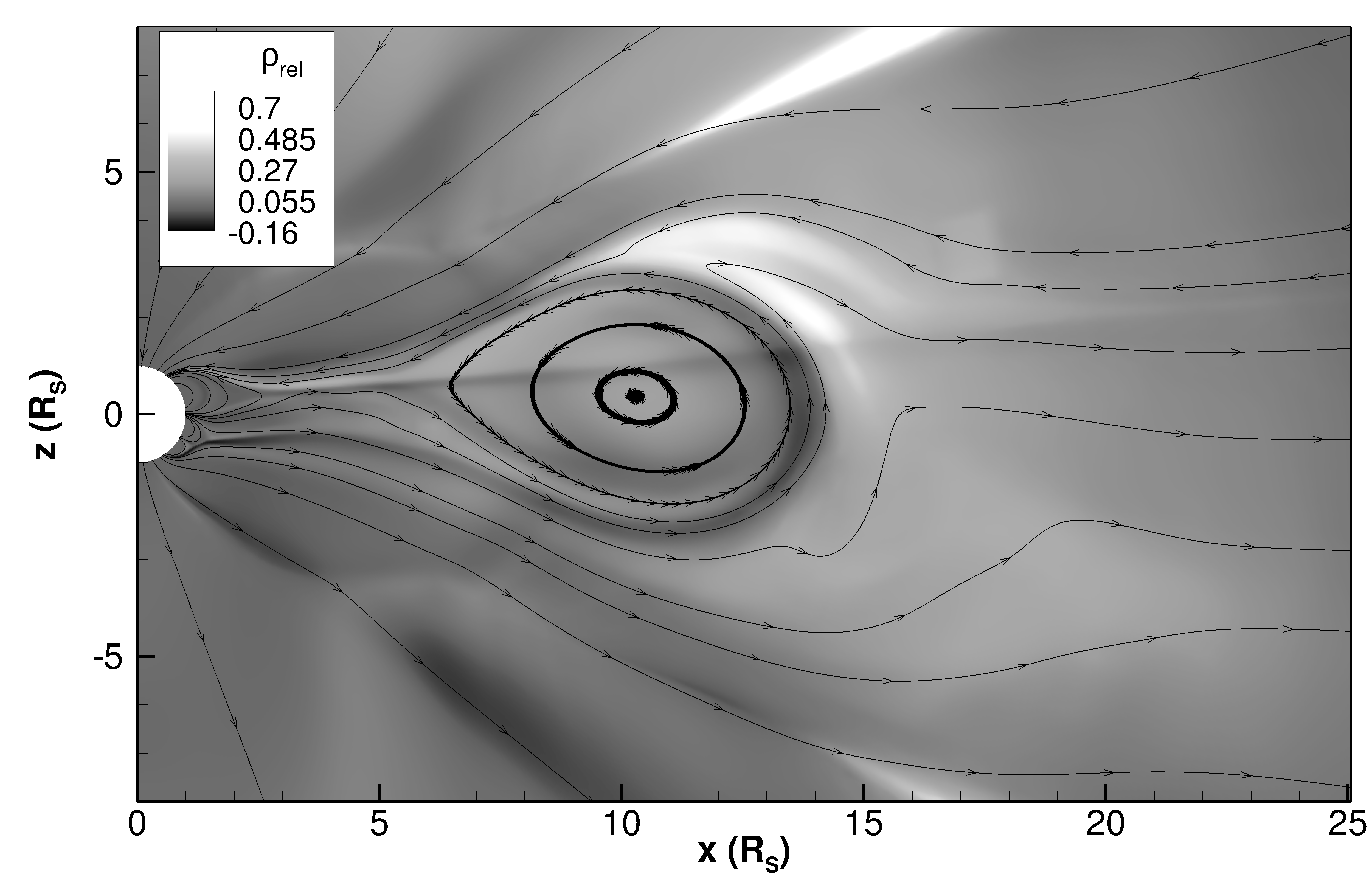}
                \end{overpic}}
        \subfloat[$\mathrm{t_{sim}= 109 h}$ \label{fig:FW_single_3}]{
                \begin{overpic}[width=0.32\linewidth]{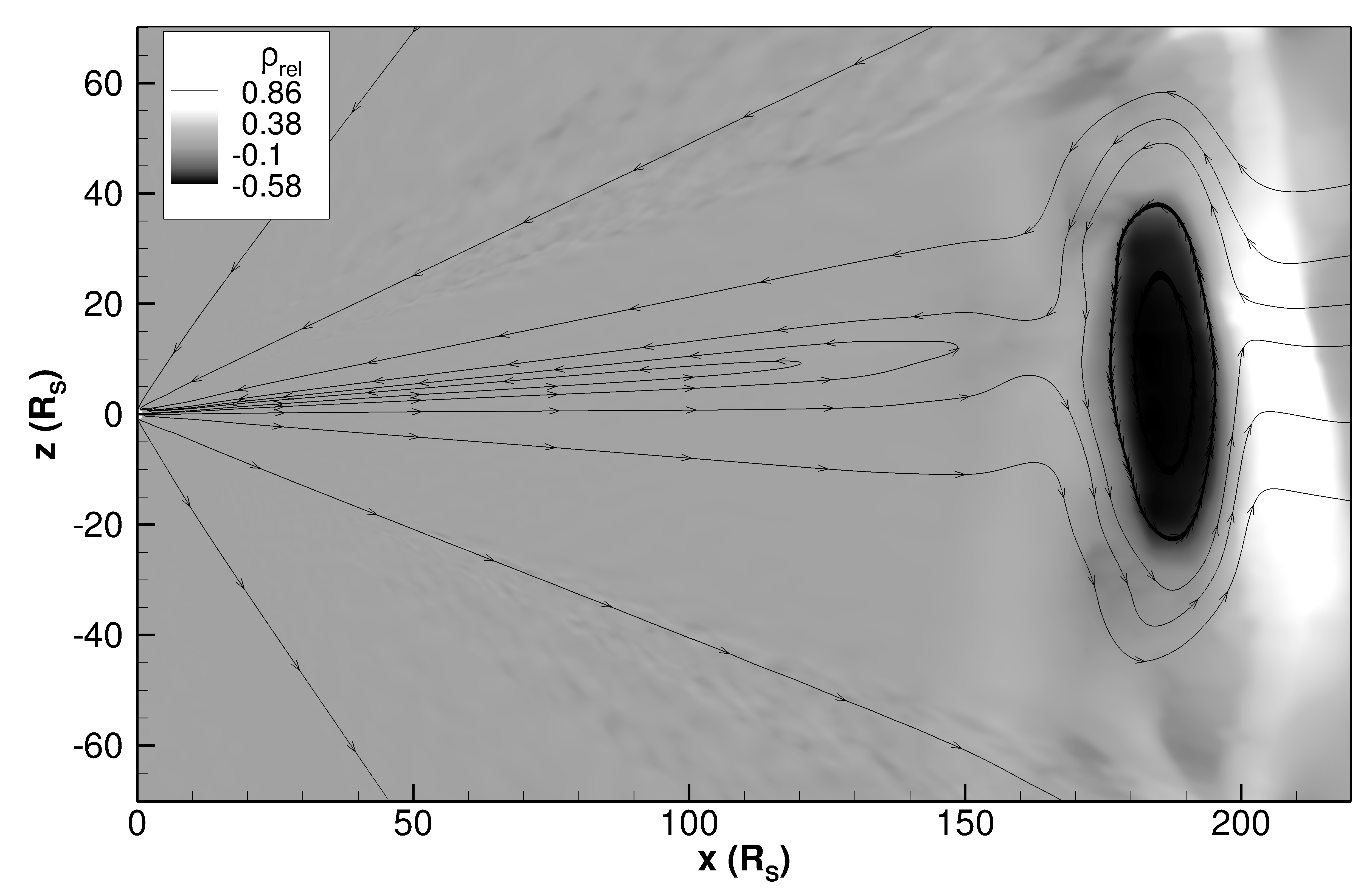}
                \end{overpic}}\\
        \vspace{-0.3cm}
        
        \subfloat[$\mathrm{t_{sim}= 13 h}$ \label{fig:FW_stealth_speed_1}]{
                \begin{overpic}[width=0.32\linewidth]{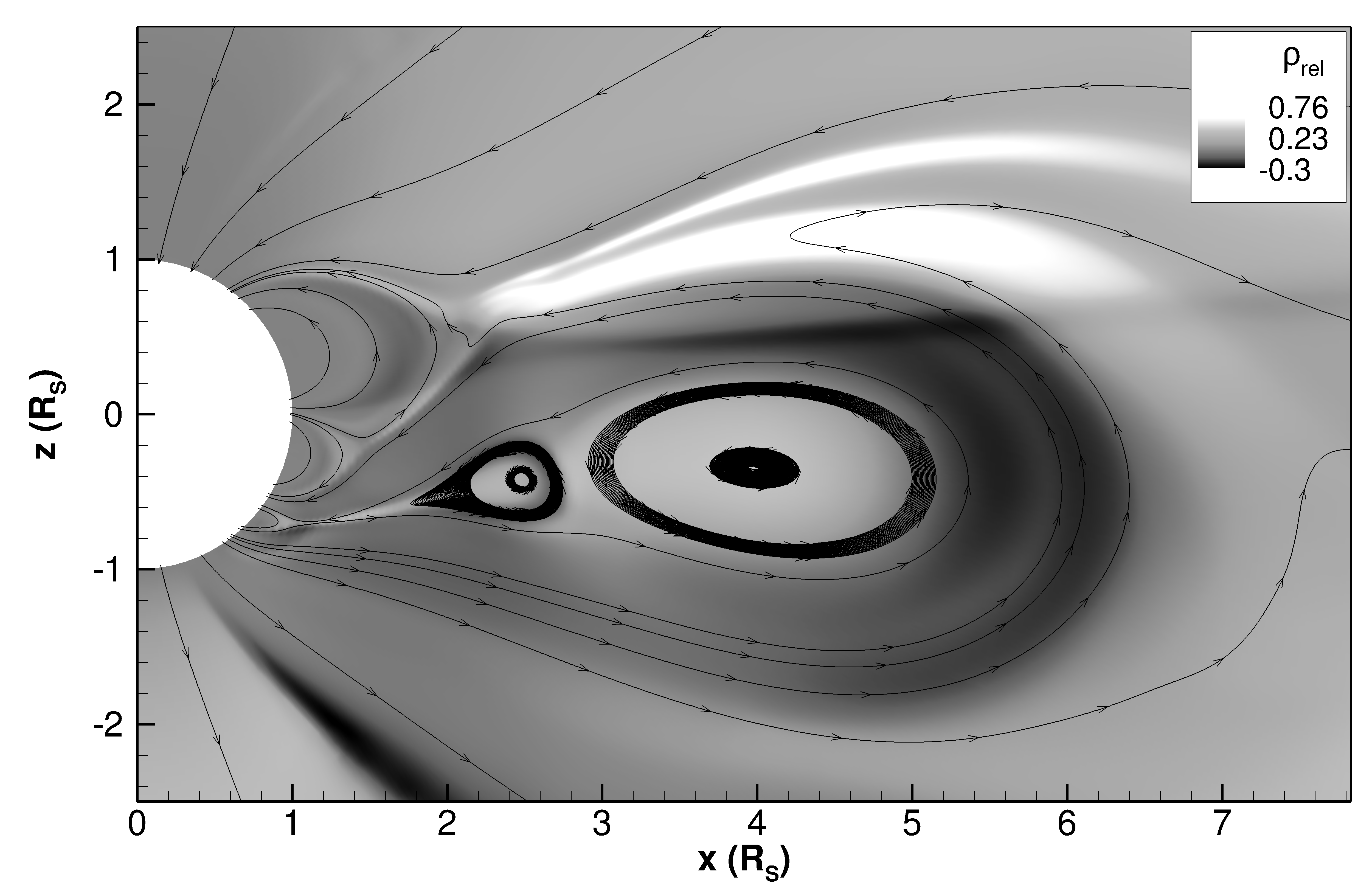}
                        \put(13,58){\tiny \textbf{FW stealth speed}}
                        \thicklines                         
                \end{overpic}}
        \subfloat[$\mathrm{t_{sim}= 22 h}$ \label{fig:FW_stealth_speed_2}]{
                \begin{overpic}[width=0.32\linewidth]{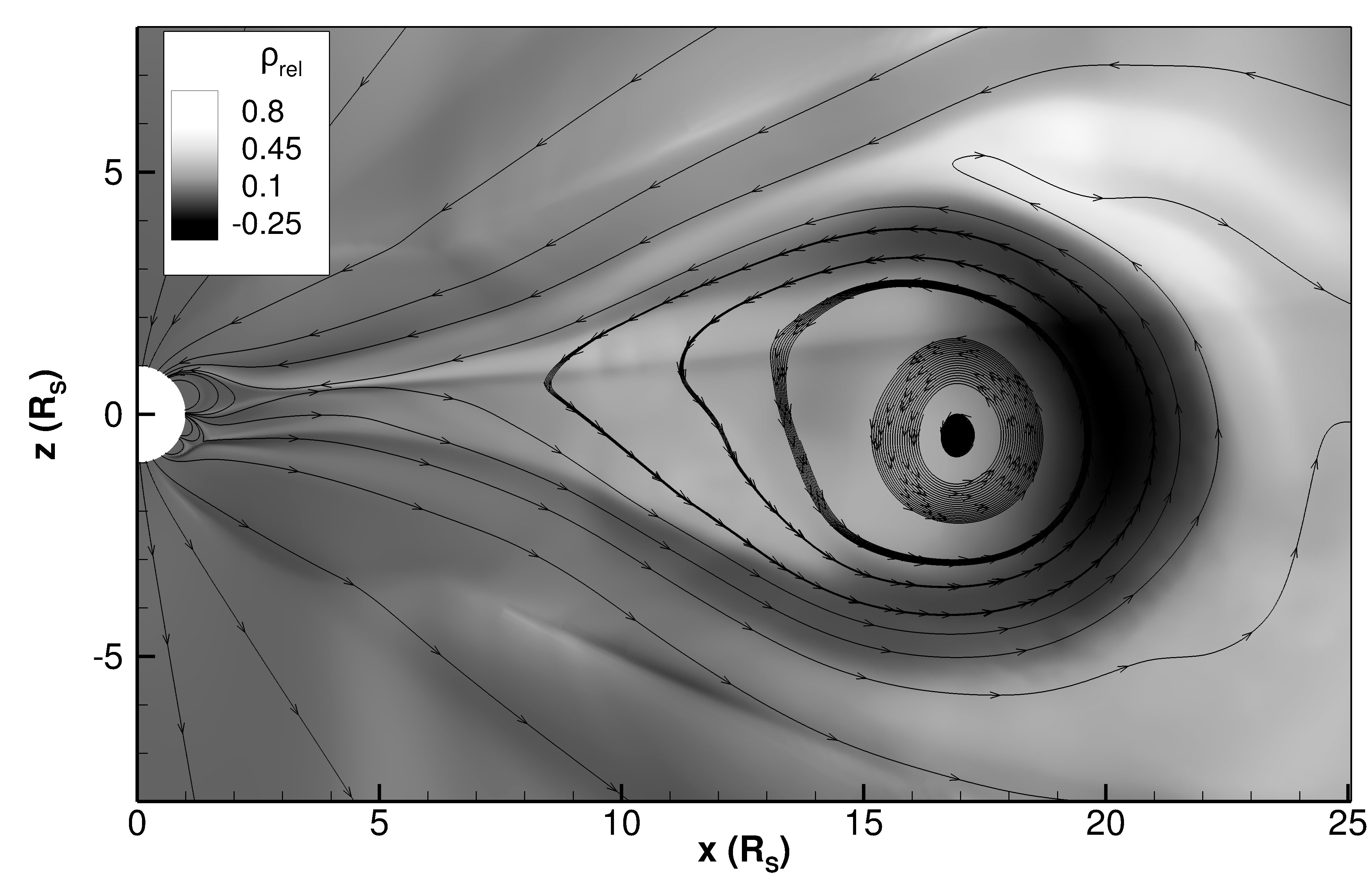}
                \end{overpic}}
        \subfloat[$\mathrm{t_{sim}= 104 h}$ \label{fig:FW_stealth_speed_3}]{
                \begin{overpic}[width=0.32\linewidth]{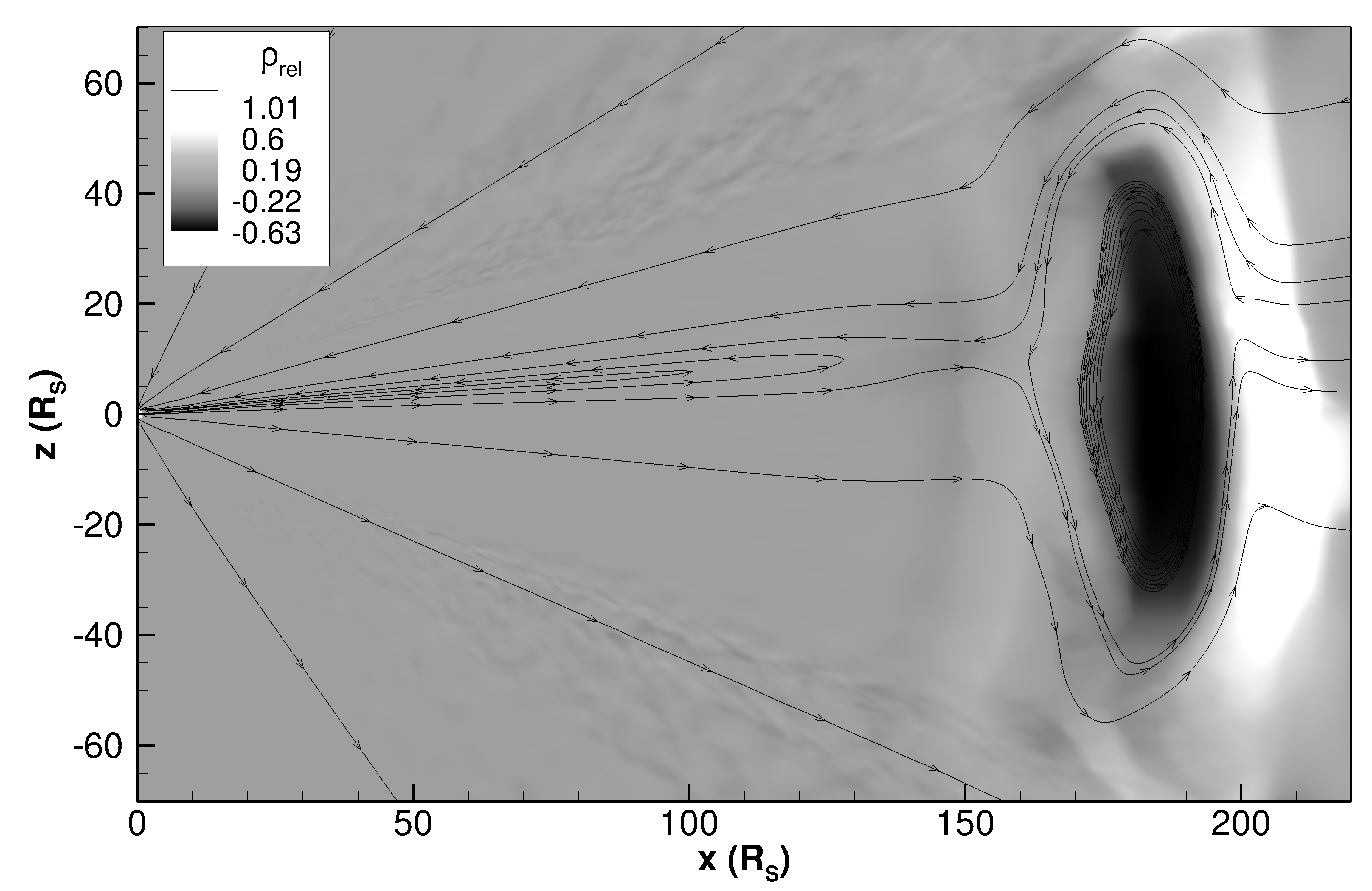}
                \end{overpic}}\\

        \caption{Simulated relative density (grey scale) and selected magnetic field lines during the eruption of all CMEs, in the case of: first row --  single eruption (ejected into the slow wind (SW)); second row -- eruption + stealth (slow wind); third row -- double eruption (slow wind); fourth row -- single eruption (faster wind); fifth row -- eruptions in the faster wind resulting from applying the same shearing speed that created a stealth ejecta in the slow wind case. The relative density is $\rho_{rel}=\frac{\rho(t)-\rho_0}{\rho_0}$, where ${\rho_0}$ is the density of the initial relaxed state before the shear. The time $\mathrm{t_{sim}}$ is counted from the start of shear. The arrows are used in the text to indicate plasma blobs.}
        \label{fig:simulations_snapshots}
\end{figure*}

In total, we performed five simulations and for each of them Fig.~\ref{fig:simulations_snapshots} shows three snapshots in time of the propagation of the erupting structures through the solar wind; these are accompanied by supplementary videos available online. For easier visual inspection, the grey scale indicates the density of the current time-step snapshot  relative to the relaxed wind state, similar to base difference images created from coronagraph images (lighter shading indicates denser plasma). In all cases, the first flux rope is formed by the applied azimuthal flow through the same physical processes. The additional $v_\phi$ firstly increases the magnetic pressure inside the southernmost arcade, expanding it and making it rise. As a consequence, an imbalance is created between the magnetic tension and magnetic pressure gradient which leads to a local compression of the magnetic field. The numerical resistivity allows the sides of the arcade to reconnect, thus creating the flux rope, which begins to erupt. The southern polar magnetic pressure deflects it towards the equator, until it starts propagating radially inside the equatorial current sheet from several solar radii onwards, depending on each scenario. The secondary eruptions (if any) differ for each case, and the five numerical simulations can be briefly described as follows (including the first CME):

\begin{enumerate}
        \item  Single eruption (slow wind): Given the low amplitude of the shearing motions, only one flux rope is formed 12 h after the start of the shear, which slowly erupts and gets deflected towards the equatorial plane, creating the main CME (Fig.~\ref{fig:simulations_snapshots}a,b). The current sheet formed in the wake of this eruption magnetically reconnects and thus five plasma blobs arise, of which only one survives during their journey to Earth and is indicated by the white arrow in Fig.~\ref{fig:simulations_snapshots}c. An extra blob forms much later during this propagation (84 h after the start of the shear) and is indicated by the black arrow in Fig.~\ref{fig:simulations_snapshots}c. The rest of the blobs magnetically reconnect with the first CME, because they are created in a depleted solar wind environment which allows them to easily catch up with their precursor. 
                
        \item  Eruption + stealth (slow wind): After the formation of the first flux rope (8 h after the shear start) triggered by shearing motions, and the ejection of the associated CME (Fig.~\ref{fig:simulations_snapshots}d), a second flux rope (stealth ejecta) is created from the reconfiguration of the coronal magnetic field and erupts in the trail of the preceding eruption (Fig.~\ref{fig:simulations_snapshots}e, indicated by the white arrow), also being deflected towards the equator. Two plasma blobs arise in this case as well (apart from the stealth ejecta) and maintain their magnetic structure until 1 AU. Similarly to the previous case, an extra blob is created 82 h after the start of the shear, and so when the CME arrives at Earth, there are three trailing flux ropes, as the stealth ejecta magnetically reconnected with the first eruption.      
        
        \item  Double eruption (slow wind): After the first CME (Fig.~\ref{fig:simulations_snapshots}g), a second flux rope also erupts (Fig.~\ref{fig:simulations_snapshots}h), both triggered by the shear applied at the boundary. During their propagation to 1 AU, the slow nature of both eruptions allows the second CME to reconnect with the first one, arriving as one entity at Earth, as seen in Fig.~\ref{fig:simulations_snapshots}i. From the five plasma blobs created, only two remain after 115 h, but the second one disappears shortly after that. Following the other two cases, another blob arises in the trailing current sheet 87 h after the start of the shear. 
                                
        \item  Single eruption (faster wind): An erupting flux rope is formed 10 h after the start of the shear, and is deflected northward into the equatorial current sheet, into which it propagates almost radially after $\approx$20 h (Fig.~\ref{fig:simulations_snapshots}j,k). Interestingly enough, there are no plasma blobs created after the CME, which arrives at Earth after $\approx$109 h as one pancaked flux rope, followed by the trailing streamer current sheet forming again in its wake (Fig.~\ref{fig:simulations_snapshots}l).    

        \item  Stealth speed (faster wind): A first flux rope is formed much faster than in the single eruption case (similarly to the slow wind scenarios), namely 7 h after the start of shear, and a second one only 12 h after the start of shearing, both being deflected northward into the equatorial current sheet (Fig.~\ref{fig:simulations_snapshots}m). The second flux rope is no longer created from the reconnection of the coronal magnetic field, but is a result of the imposed shearing motions. Their faster formation and eruption speeds led to a much earlier merging between the two CMEs, propagating as one entity as of 19 h from the start of shear (Fig.~\ref{fig:simulations_snapshots}n). Their quick reconnection created a very similar pancaked flux rope to that of the single eruption case at 1 AU (Fig.~\ref{fig:simulations_snapshots}o), but not the same magnitude of magnetic field, as will be shown in the following two sections (\ref{sec:insitu} and \ref{sec:geoeffectiveness}). Interestingly, in this case there were also no plasma blobs created, making their formation dependent on the initial magnetic configuration rather than the magnitude of the shearing speed.    
                        
\end{enumerate}

The first two columns of Fig.~\ref{fig:simulations_snapshots} are snapshots of simulations taken at the same time in order to show a comparison between the formation and eruption of the CMEs created in the two different background winds. We note the difference in density, for instance in the single eruption cases. The flux rope in the slow wind is filled with plasma, whereas in the faster wind, the front is more compressed due to the higher speed, and therefore is denser than the core. However, this ratio  is not kept during the propagation to 1 AU, probably due to the drag forces exerted by the background wind and by the different initial densities as a result of higher temperatures.


    \section{Comparison with in situ signatures} \label{sec:insitu} 

Simulations should show some agreement with observations if they are to help us understand the physics of actual events. Once the simulation is realistic, it is interesting to investigate how different variables influence the eruption itself and the propagation through the interplanetary space, because one particular event is not representative of the multitude of eruptions and solar parameters. Fortuitously, the MCME described in Section \ref{sec:observations} erupted on 21-22 Sept 2009, and encountered spacecraft at Mercury and Earth, allowing us to compare our simulation results with in situ observations at these locations. \par 
The comparison close to the Sun between these two simulations and the event was performed in paper I, and here we are focusing on the interplanetary part and in situ signatures. We took 1D slices in our 2.5D simulations on the equatorial plane and extracted the data at 75 R$_{\sun}$ and 215 R$_{\sun}$ in order to compare with measured values at Mercury and at Earth. As this eruption takes place at the time of the equinox, the equatorial and ecliptic plane coincide, and so no latitudinal correction is needed. On 23 Sept 2009 (when the CME arrived at Mercury), the MESSENGER spacecraft was at an angle of only $\approx$11$\degr$ westward from the Sun--Earth line at a distance of $\approx$75 R$_{\sun}$ from the Sun and not in Mercury's magnetosphere. This makes the onboard Magnetometer \citep{mag_messenger} instrument ideal for in situ magnetic field data comparison with our simulations. The in situ values at Earth (L1) were taken from the OMNI database, which combines data measured by the WIND \citep{SWE_WIND_spacecraft, MFI_WIND_spacecraft} and ACE \citep{ACE_spacecraft} spacecraft to provide near-continuous solar wind observations. The cadence of the data used was 1 min for MESSENGER and 5 min for OMNI, both averaged over 20 minute intervals in order to reduce fluctuations that are on much smaller scales than those produced in the simulation, but also to ease inspection of the structures of interest. For a proper comparison with the in situ data, we also averaged our simulation values over 20 minutes, because even though they lack the small-scale structures, the amplitude of the investigated flux ropes are also affected by the averaging, and we wish to be consistent in our study. We mention that the simulation time has been matched with the observed time at the moment when the dark cavity of the first CME was still just within the COR1-B field of view, at 23:00 on 21 Sept 2009. From then on, we no longer adjust the simulation time. This process was required because the simulation time starts counting at the start of shear when $\mathrm{t_{sim}=0\ h}$, and it needs to be given a real date, otherwise the data cannot be compared.         

\begin{figure}[h!]
        \centering
                \begin{overpic}[width=1\linewidth]{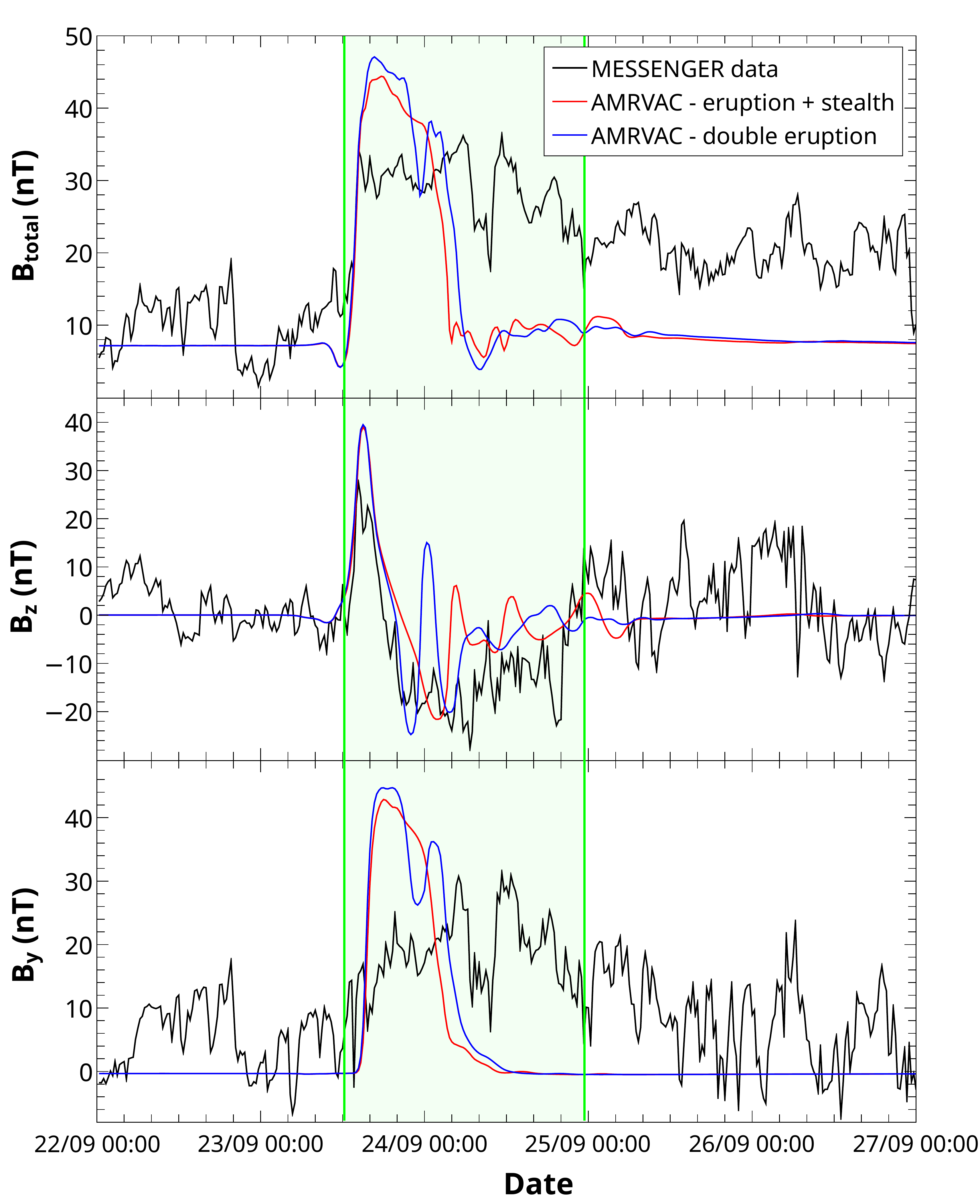}  
                        \put(0,0){\color{gray} \thicklines \dashline[30]{1.5}(40,8)(40,96)}  
                        \put(0,0){\color{cyan} \thicklines \dashline[30]{1.5}(34.9,8)(34.9,96)}
                \end{overpic}
        \caption{In situ measured components of the magnetic field taken by MESSENGER (black line), and simulation data in the eruption + stealth (red line) and double eruption (blue line) cases, propagated in the slow wind and taken at 75 R$_{\sun}$. The green highlighted area approximately delimits the observed ICME. The cyan and grey dashed lines indicate the separation between the two flux ropes in the simulated double eruption case, and observed data, respectively. The dates on the X axis are from the year 2009.}
        \label{fig:messenger_amrvac} 
        \vspace{-0.4cm}         
\end{figure}

\begin{figure}[h!]
        \centering
        \resizebox{\hsize}{!}{\includegraphics{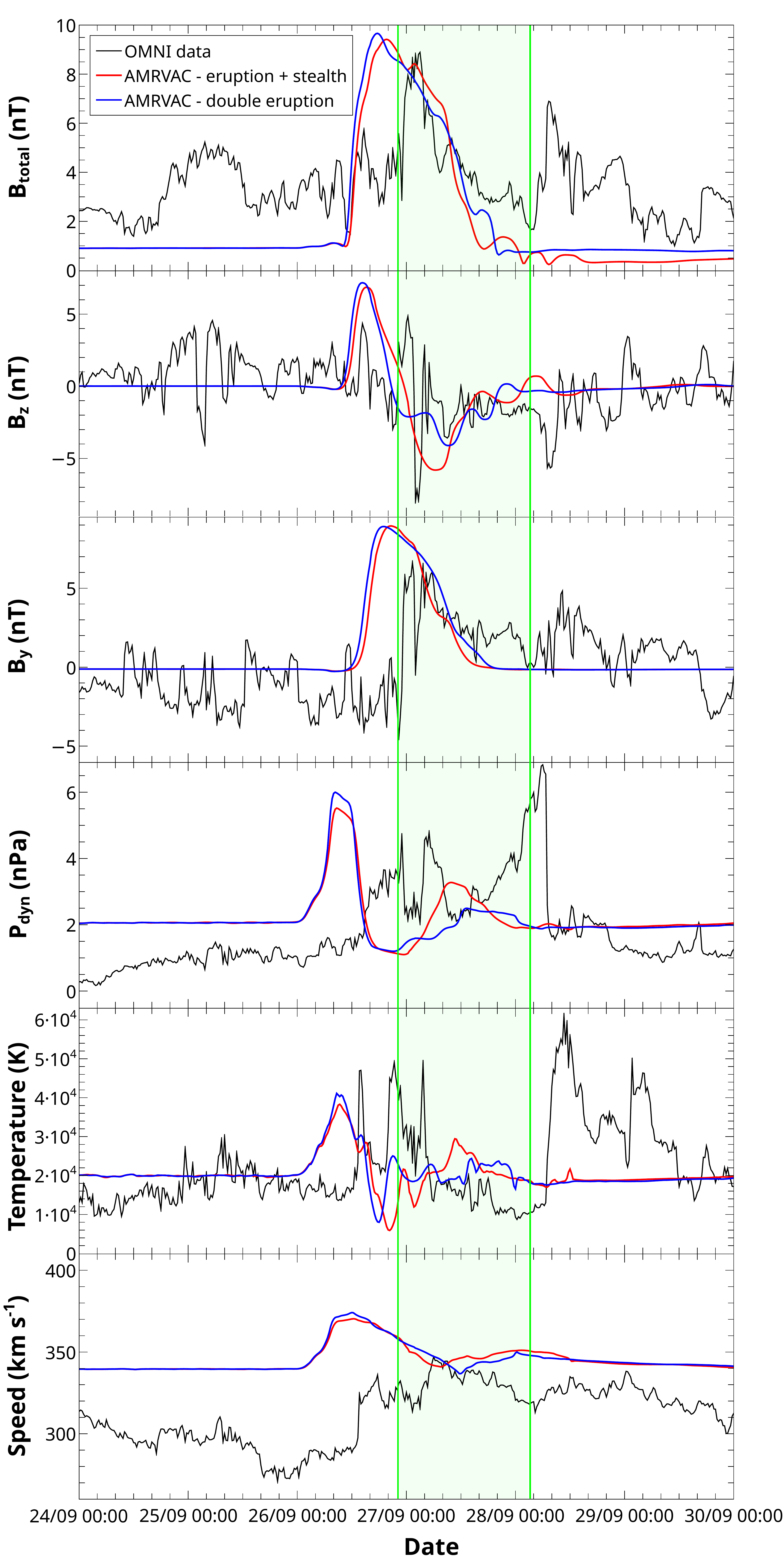}}
        \caption{In situ measured values of the magnetic field components, dynamic pressure, temperature, and speed of the solar wind (top to bottom) taken from the OMNI database (black line). Also  plotted are simulation data in the eruption + stealth (red line) and double eruption (blue line) cases propagated in the slow wind and taken at 1 AU. The green highlighted area approximately delimits the ICME. The dates on the X axis are from 2009.}
        \label{fig:omni_amrvac}
        \vspace{-0.4cm}         
\end{figure}

In Fig.~\ref{fig:messenger_amrvac}, the ICME can be distinguished in MESSENGER data by the increase in total magnetic field and $B_y$, as well as the fairly smooth rotation in the $B_z$ component, criteria based on which the green area has been delimited. The simulation data show similar trends for both cases, eruption + stealth and double eruption, as well as a perfect arrival time (at Mercury). The second flux rope can still be differentiated from the first one in the observed data, and it starts at the dip in the total magnetic field and in the $B_y$ component on 24 Sept at $\approx$08:00, delimited by the grey dashed line. The simulations also show a second flux rope, which is much more pronounced in the double eruption case (separation indicated by the cyan dashed line), because the second CME in this scenario is wider and has a stronger magnetic field than the stealth CME. However, the main difference between MESSENGER and AMRVAC data is that the observed signatures of total magnetic field and the $B_y$ component last almost twice as long.\par 
During the propagation until 1 AU, the influence of the real solar wind exhibits a larger impact than closer to the Sun and its drag force is dominant as compared to other forces, which distorts the signatures. This effect can be seen in the magnetic field components measured near Earth and shown by the black lines in Fig.~\ref{fig:omni_amrvac} in the top three panels. The arrival time at Earth in simulations is also affected by the faster background solar wind speed at the equator (340 km s$^{-1}$), as compared to the very localised minimum speed recorded in the northward-shifted current sheet (330.6 km s$^{-1}$, Fig.~\ref{fig:speeds_1AU}). We noticed in AMRVAC data the well-known and thoroughly studied pancaking effect of the frontal flux rope, but also the merging of both secondary flux ropes with the first one, as they are ejected into a depleted solar wind from the passage of the first CME. Even though the first two flux ropes are reconnected, the signature of the previously present second CME is stronger in the double eruption case than in the stealth ejecta, as you can see in the evolution of the $B_z$ component  (blue and red lines, second panel in Fig.~\ref{fig:omni_amrvac}). After $B_z$ turns negative, it increases again but no longer changes sign, which also affects the minimum value of this component. As the CME arrives at Earth with a positive frontal $B_z$, the negative trailing magnetic field is diminished due to the reconnection with the following flux ropes, leading to an increased (closer to zero) value of the negative part. Some plasma blobs can still be distinguished by the small oscillations in the trail of the ICMEs.\par 
The compressed front of the first CME evident in observations is reproduced well in simulations, as seen in the first peak of the dynamic pressure ($P_{dyn}$ panel in Fig.~\ref{fig:omni_amrvac}), which is defined as half the mass density multiplied by the square of the total speed of the solar wind. The observed proton temperature also presents an increase in the frontal part, followed by the usual lower values inside the flux rope, a signature observed in a large percentage of ICMEs \citep{low_temp_richardson, in-situ_signatures_ICMEs}. The simulations do not show such an extended interval of low temperatures due to the trailing plasma blobs, but they do exhibit the decrease in temperature. The final panel of Fig.~\ref{fig:omni_amrvac} describes the expansion of the magnetic cloud through the faster front and lower speed of the tail. \par       
The arrival time of the fronts of the CMEs  at 1 AU is very similar for both simulated cases (double eruption and eruption + stealth), with a difference of only 20 min in magnetic field components, which is given by the slightly different eruption times. This is expected because the different eruption times of the CMEs are a consequence of the extremely small variation in the triggering shearing speed when there is a similar background wind and a higher amplitude of $v_\phi$ resulting in a faster flux rope formation and ejection. The difference in arrival time between the simulations and observations is $\approx$ 10 hours, which is introduced by the differently modelled background solar wind speed, but seems to be a typical error for such simulations; see for example \citet{mays_simulations_cmes}. Nevertheless, this is a surprisingly good result considering that the wind was 2.5D simulated to  only approximately match the 1 AU in situ measured speed. Furthermore, all the extracted parameters qualitatively fit the observed variables.     

\begin{figure}[h!]
        \centering
        \resizebox{\hsize}{!}{\includegraphics{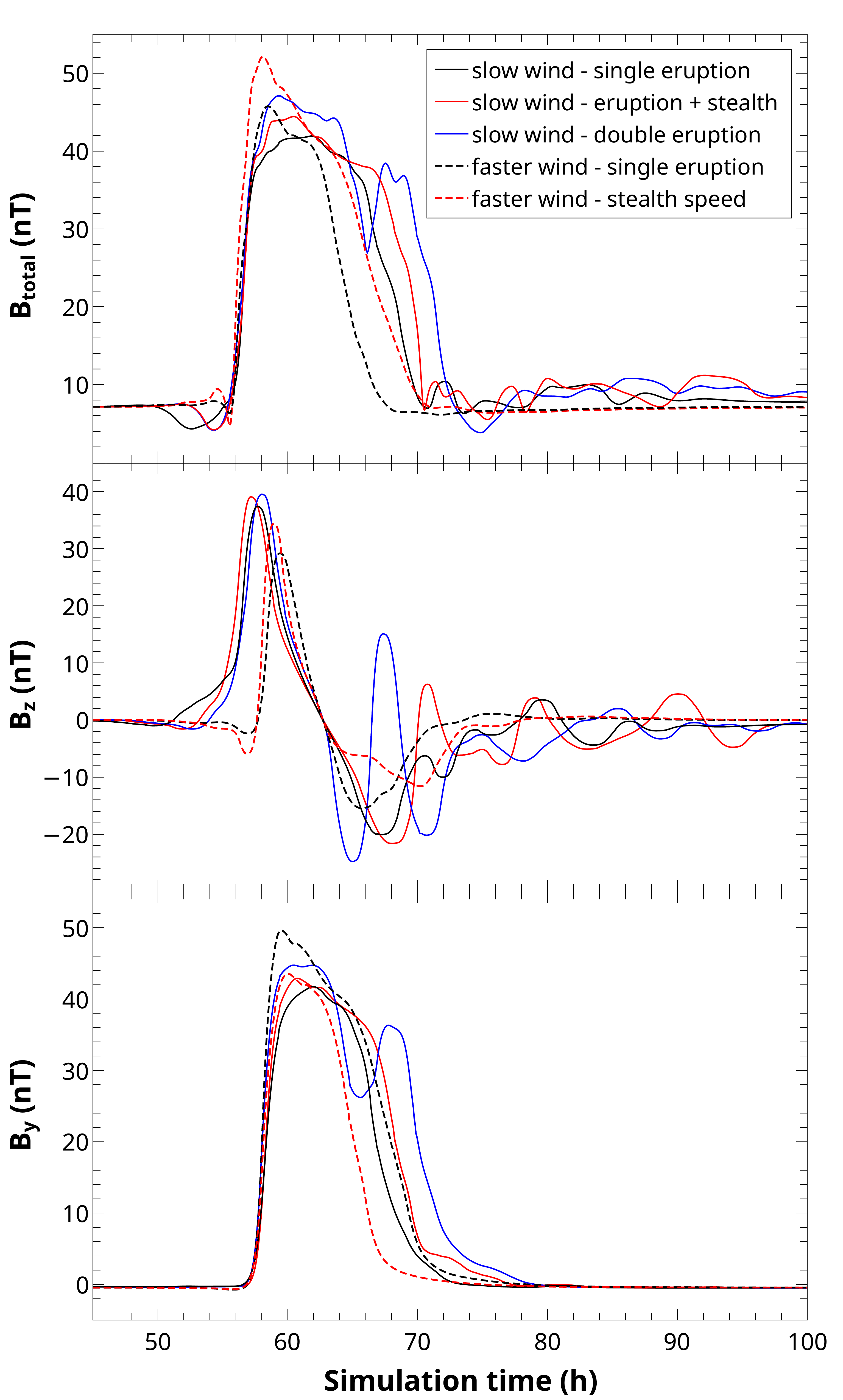}}
        \caption{Magnetic field components simulated in all five cases, taken on the equatorial plane at 75 R$_{\sun}$. All simulation data are shifted and aligned with the last arriving CME, which occurs in the slow wind, single eruption case. $B_{total}$ and $B_y$ have their fronts aligned, whereas $B_z$ is aligned at the point where the values change sign inside the first main CME.}
        \label{fig:sims_75Rs_signatures}
        \vspace{-0.4cm}         
\end{figure}

\begin{figure}[h!]
        \centering
        \resizebox{\hsize}{!}{\includegraphics{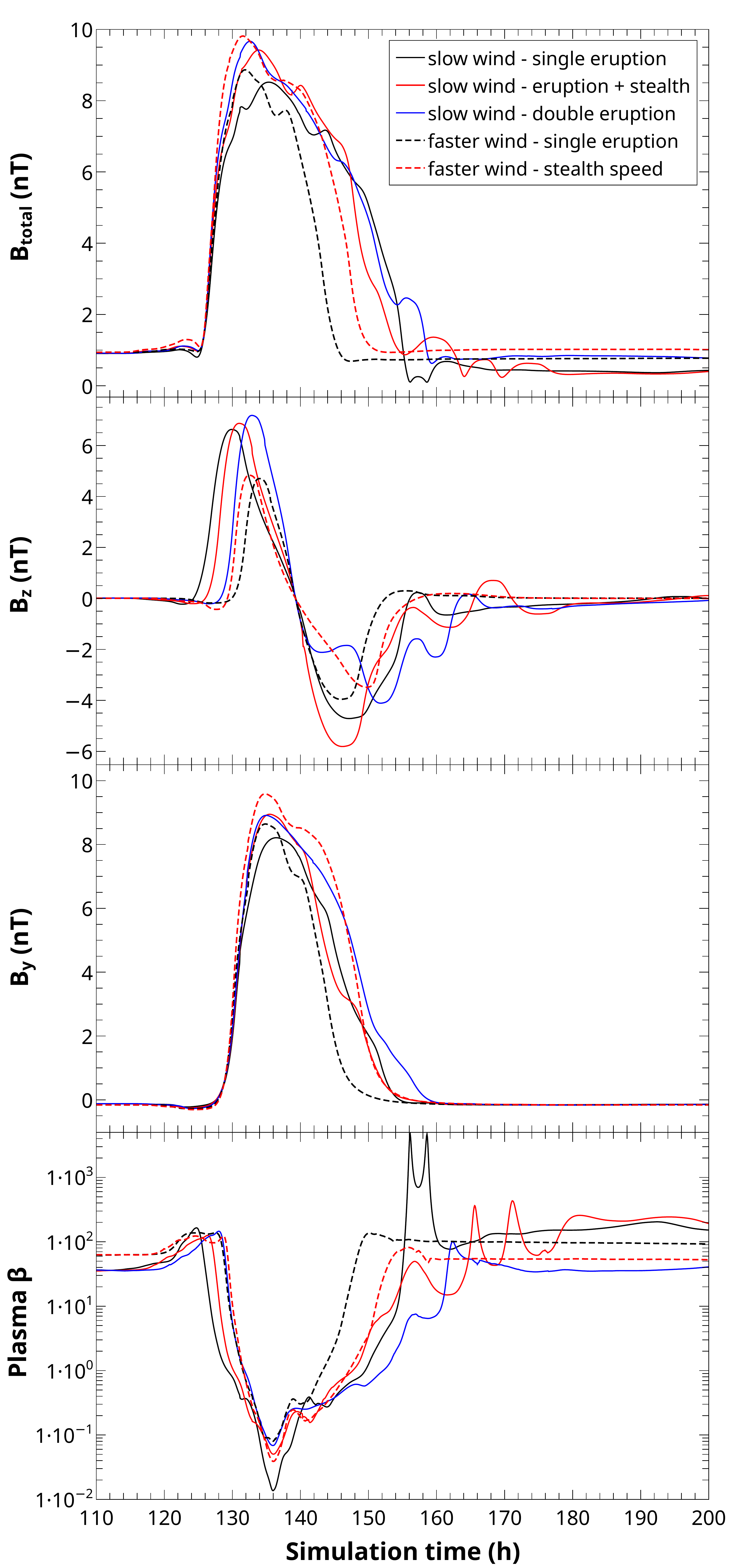}}
        \caption{Magnetic field components simulated in all five cases, taken on the equatorial plane at 1 AU. The magnetic field components are overlapped in the same way as in Fig.~\ref{fig:sims_75Rs_signatures}, and we overlapped the minima of the plasma beta curves with the curve of the slow wind, single eruption case.}
        \label{fig:sims_1AU_signatures}
        \vspace{-0.4cm}         
\end{figure}

The fact that the two discussed simulations show good correlation with observations suggests that the initial setup is realistic enough to further investigate other eruptive scenarios. We are also interested in  analysing how the second CME influences its precursor, and to do so we simulated a single erupting flux rope, as previously described in Section \ref{sec:eruptions}. Furthermore, we investigated how a faster background solar wind would influence the propagation of such shear-induced CMEs, and we focus in the next part of this section on the five simulations as an independent result, rather than comparing them with the observations. The five discussed scenarios are illustrated in Section \ref{sec:eruptions}.\par  
For each of these simulations, the magnetic field components were extracted in the equatorial plane at 75 R$_{\sun}$ (Fig.~\ref{fig:sims_75Rs_signatures}) and at 215 R$_{\sun}$ (1 AU, Fig.~\ref{fig:sims_1AU_signatures}). For an easier comparison, the curves have been shifted in time and aligned with those of the last arriving CME, which is always the single eruption inserted into the slow solar wind. In both figures, $B_{total}$ and $B_y$ have their fronts aligned, the $B_z$ curves are aligned at the point where the values change sign inside the first flux rope, and in Fig.~\ref{fig:sims_1AU_signatures} plasma beta curves are aligned at their minima. Plasma beta is defined as the ratio between plasma pressure and magnetic pressure. The process of aligning the curves leads to a loss of the information of arrival time at the respective distances, and so we note those explicitly in Table \ref{table:arrival_times}. 

\begin{table}[h!]
        \centering
        \caption[]{Arrival times of the total magnetic field dip (followed by enhancement) of all simulated CMEs at 75 R$_{\sun}$ and at 215 R$_{\sun}$. Here, SW = slow wind and FW = faster wind.}
        \begin{tabular}{c c c} 
                \label{table:arrival_times}
                Eruption type             & Arr. time at 75 R$_{\sun}$   &  Arr. time at 215 R$_{\sun}$ \\
                \hline          
                \noalign{\smallskip}
                        {Single er. SW}   & 52.6 h       & 124.8 h \\
                        {Stealth er. SW}  & 48.13 h      & 119.2 h \\
                    {Double er. SW}   & 48.13 h       & 118.53 h \\                                                                
                \noalign{\smallskip}                          
                \hline
                \noalign{\smallskip}
                        {Single er. FW}     & 49.4 h   &  113.93 h  \\
                        {Stealth speed FW}  & 45.4 h      &  109.67 h \\
                \noalign{\smallskip}
                \hline 
        \end{tabular}
\end{table}          

Table \ref{table:arrival_times} was created using the arrival times of the minimum of the small dip that is created in front of the total magnetic field due to turbulence and compression of the equatorial current sheet ahead of the CME, and therefore this dip might not correlate precisely with the front of CMEs. We used this feature because of its precision, as compared to the magnetic field enhancement which would require a subjective choice of starting time. However, the time difference is insignificant, namely of the order of tens of minutes. \par   
The total magnetic field taken at 75 R$_{\sun}$ presents some major differences between the eruptions inserted in the slow and faster wind, namely in the widths of the ICMEs and the presence or absence of following plasma blobs. The double eruption case exhibits the widest ICME structure due to the yet-to-be-reconnected second CME, whereas the single eruption in the faster wind is the shortest in time. However, the stealth speed case has the largest total magnitude of magnetic field, which is also the case at 1 AU (top panel of Fig.~\ref{fig:sims_1AU_signatures}). The weakest magnetic field is found in the single eruption case of the slow wind, both at Mercury and Earth. The signatures seem to be sharper in the faster wind cases than in the slow wind ones, which is probably related to the compression of the front attributed to the higher speed of CMEs and denser background solar wind. The second flux rope is still present in both double eruption and stealth ejecta cases, but can only be clearly distinguished in $B_{total}$ and $B_y$ in the first case. On the other hand, $B_z$ (middle panel of Fig.~\ref{fig:sims_75Rs_signatures}) shows a larger variation and more clearly indicates the flux ropes erupting in the trail of the first CME in the slow wind cases.\par 
The plasma blobs appear to be well defined in the $B_z$ component, but in $B_y$ there is almost no difference in the tail between the slow and faster wind cases, and so the blobs are not visible in this component either at 75 R$_{\sun}$ (bottom panel of Fig.~\ref{fig:sims_75Rs_signatures}) or at 215 R$_{\sun}$ (third panel of Fig.~\ref{fig:sims_1AU_signatures}).\\
We propose a possible explanation for the absence of plasma blobs in the faster wind scenarios. The first CME is deflected towards the equator in all simulations, thereby compressing the northern arcade. In the faster wind configuration, this structure does not extend as far as in the slow wind, and is not affected as much by the CME; therefore, it does not relax to the initial state, because it was not as perturbed. In the slow wind cases, the height of the northern arcade is greatly reduced during the deflection of the first CME, and in the process of returning to its initial state, it compresses the current sheet, inducing reconnection and creating plasma blobs just above the streamer cusp. This suggests the blobs may be indicators of more closed magnetic structures in the source region before the eruption.\par    
It is interesting to note that the negative $B_z$ components are smaller for the faster solar wind cases than for the slower wind scenarios, and that the weakest negative $B_z$ is found in the stealth speed simulation and is due to the second shear-triggered CME that reconnects with the trailing part of the first one. Comparing this to the total magnetic field, we can see that the contribution of $B_z$ is not as significant as the other components, in particular $B_y$. In this 2D visualisation, the $B_y$ component is equivalent to $B_\phi$, and is the axial or toroidal magnetic field along the centre of the flux rope, while $B_z$ is the azimuthal field (twist or poloidal component). Therefore, $B_y$ is a consequence of the shear component of the magnetic field introduced by the $v_\phi$ boundary motions, whereas $B_z$ originates from the reconnection of overlying streamer flux with itself, in the process of creating the twisted flux rope. The overlying streamer flux is larger in scale than the low-lying shear component, and therefore has lower field magnitude, which is also true in many flux rope models used to fit in situ data \citep[e.g.][]{lepping_magnetic_cloud_structure, burlaga_force_free_field}. This also explains the presence of the second peak in $B_y$ in the double eruption case, because the second CME is triggered by the shearing motions and its flux rope is created close to the inner boundary. This characteristic of showing an axial field magnitude larger than the twist field magnitude  also propagates to 1 AU, and is an important factor in estimating the geoeffectiveness of the simulated CMEs in Section \ref{sec:geoeffectiveness}. \par
The interaction with the background solar wind until 1 AU and the latitudinal expansion and radial compression of the flux ropes led to several changes in the ICME signatures. The $B_z$ components in the faster wind cases are more symmetric with respect to the centre of their flux ropes, but they still carry the lowest absolute values of all the simulations, as can be seen in the second panel of Fig.~\ref{fig:sims_1AU_signatures}. The last panel shows the ICMEs lengths through the plasma beta parameter, which usually shows a decrease inside the magnetic cloud due to the enhanced magnetic field \citep{lepping_magnetic_clouds}, and it is clearly seen that the single eruption inserted in the faster wind is the most compressed and short flux rope, followed by the stealth speed case. The slow wind simulations present larger flux ropes, as well as plasma blobs following the main eruptions, distinguished by the peaks and oscillations in plasma beta. Some curves might present stronger peaks than others, but this only indicates that the selected trajectory through the simulation runs close to the centre of these flux ropes, or almost through the null points between flux ropes, where there is approximately zero magnetic field. The well-known decrease in plasma beta inside the main ICME \citep{manchester_cmes_interaction} occurs in all simulations. \par


    \section{Geoeffectiveness} \label{sec:geoeffectiveness} 
 
The last topic of this paper is our investigation of the potential impact of the simulated CMEs on Earth's magnetosphere. One way of quantifying the geoeffectiveness of CMEs is through the $Dst$ index, which measures changes in the horizontal component of the magnetic field at ground level \citep{sugiura_dst}, and can be predicted from solar wind parameters using various empirical models. Depending on the number of mechanisms and parameters taken into consideration, these models can be rather rudimentary \citep{burton_dst_1975}, or fairly comprehensive, as in the algorithm of \citet{temerin&li_2002,temerin&li_2006}. We chose to compute our simulated $Dst$ using an intermediate model introduced by \citet{dst_computation}, which achieves a sufficient level of accuracy when compared to the measured values of $Dst$ taken from the World Data Center at Kyoto University\footnote{http://wdc.kugi.kyoto-u.ac.jp/dstdir/}. The three main parameters that contribute, in this description, to the geoeffectiveness of interplanetary structures are the speed of the incoming solar wind, the N-S component of the magnetic field, and the dynamic pressure \citep{nandita_sources_of_GSs}. Depending on the sign of $B_z$, the product between the first two variables provides the $y$ component of the solar wind convective electric field ($VB_S$) in geocentric solar magnetospheric (GSM) coordinates \citep{thompson_coordinates}, as follows:                  

\vspace{-0.2cm}
\begin{equation} \label{eq:VBs}
\hspace{0cm} VB_S \mathrm{[mV\ m^{-1}]}=  
        \begin{cases}
                |VB_z| , & B_z<0\\      
        \hspace{0.55cm}0 ,      & B_z \geq 0.    
        \end{cases}     
\end{equation}

This electric field uniquely determines the rate of ring current injection function $Q$:  

\vspace{-0.3cm}
\begin{equation}
\hspace{0cm} Q \mathrm{[nT\ h^{-1}]} =  
\begin{cases}
a(VB_S-E_c) , &VB_S > E_c\\     
\hspace{1.62cm}0,       & VB_S \leq E_c 
\end{cases}     
,
\end{equation}

\noindent where $a=-4.4$ nT m $\mathrm{h^{-1} mV^{-1}}$ and $E_c=0.49$ mV $\mathrm{m^{-1}}$.
The corrected $Dst$ index ($Dst^*$) is defined as:

\vspace{-0.2cm}
\begin{equation} \label{eq:dst_corrected}
Dst^* = Dst-b \sqrt{P_{dyn}}+c,
\vspace{-0.2cm}
\end{equation}
\noindent  where $b$ is a pressure correction term of value $7.26$ $\mathrm{nT (nPa)^{-1/2}}$, and $c$ is a constant that accounts for the quiet day currents.    
The approximate evolution in time of the corrected $Dst$ index is given by the discrete form of the Burton equation:

\vspace{-0.3cm}
\begin{equation}
Dst^*(t+\Delta t) \approx Dst^*(t) + \left[Q(t)-\frac{Dst^*(t)}{\tau}\right] \Delta t,
\end{equation}

\noindent where $\tau$ is the recovery storm time and was considered by \citet{dst_computation} to show the best fit to data in the following form:

\vspace{-0.5cm}
\begin{equation}
\tau \mathrm{(hours)}=2.4exp\left[ \frac{9.74}{4.69+VB_S}\right].
\end{equation}

The change in the dynamic pressure term determines the sudden commencement (the initial positive excursion in $Dst$ at the start of the storm) amplitude, as $Dst^*(t+\Delta t)$ should increase because of the difference in this correction term. \par
\begin{figure}[h!]
        \centering
        \resizebox{\hsize}{!}{\includegraphics{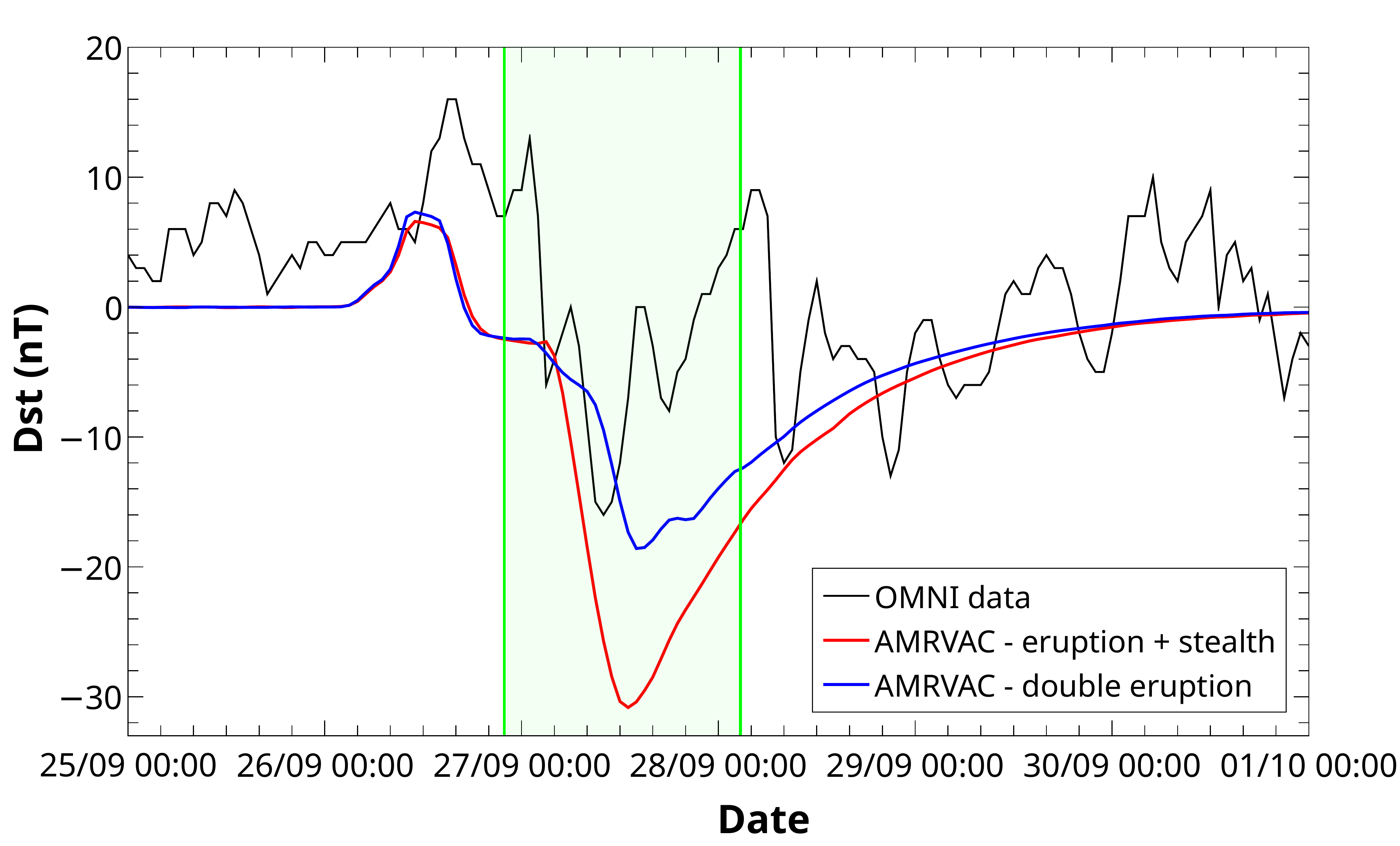}}
        \caption{Comparison of hourly $Dst$ index from observed database (black line) with modelled $Dst$ using simulation data, in the eruption + stealth (red line) and double eruption (blue line) cases, propagated into the slow solar wind. The green highlighted area approximately delimits the ICME arrival. The dates on the X axis are from the year 2009.}
        \label{fig:dst_omni_amrvac}
        \vspace{-0.3cm}   
\end{figure}

In the case of our simulations, the steady background solar wind should produce no change in the $Dst$ index, such that $Dst^*(t+\Delta t) = Dst^*(t)$, and the initial value before the start of the ICME effect should also be zero. From these conditions, we find two separate $c$ constants (from Eq.~\ref{eq:dst_corrected}) for our simulations, with values of 10.44 nT for the slow wind and 15.77 nT for the faster wind. These values differ from the ones computed by \citet{burton_dst_1975} and \citet{dst_computation}, of namely 20 nT and 11 nT, respectively, because of the different observed datasets used in their studies. This discrepancy is not unexpected because our winds are uniform in time whereas the in situ measured data reveal a wind that varies greatly over time.\\ 
We used the simulated data (speed, $B_z$ and dynamic pressure) in all of the above equations and computed a predicted $Dst$ index that our CMEs would create. Figure~\ref{fig:dst_omni_amrvac} shows a comparison between the OMNI hourly $Dst$ index and the hourly averaged $Dst$ computed from simulation data extracted every 4 min, in the slow wind cases of eruption + stealth and double eruption. Given the weakness of the observed CMEs (slow speed and only brief intervals of modest southward $B_z$ (Fig. \ref{fig:omni_amrvac})), the corresponding ICME did not create a geomagnetic storm \citep[$Dst$ $\leq -30$ nT,][]{geomag_storm_classification, gonzalez_geomagnetic_storm}, reaching a minimum measured $Dst$ of only -16 nT. As mentioned in Sect. \ref{sec:insitu}, the simulated trailing flux ropes reconnected with the tail of the first CME, and therefore the negative $B_z$ component in the double eruption case was not as strong as in the eruption + stealth scenario (see Fig.~\ref{fig:omni_amrvac} and \ref{fig:sims_1AU_signatures}). It is well known that a strong decrease in the negative $B_z$ results in a stronger $Dst$ \citep[e.g.][]{tsurutani_Bz_majorGS}. In accordance with this correlation, the double eruption case resulted in a less geoeffective ICME than the eruption + stealth scenario, with $Dst$ values resembling those recorded at Earth (see Fig.~\ref{fig:dst_omni_amrvac}).\\
Most previous studies on this topic concluded that more intense geomagnetic storms were associated with multiple interacting CMEs, rather than single CMEs \citep[][and many others]{compound_stream_majorGS_burlaga, zhang_sources_of_majorGS, gopalswamy_HCME_geoeffectiveness}. Surprisingly, our simulations highlight that the presence of interacting CMEs does not necessarily imply strong geoeffectiveness, at least in the case of slow CMEs. In our simulations, the eruption + stealth scenario is predicted to create a weak geomagnetic storm (minimum $Dst=-30.83$ nT), whereas the double eruption is predicted to be even less geoeffective, with minimum $Dst=-18.6$ nT. However, this does not necessarily contradict the previous studies, quite the contrary. Their direction of analysis was mainly starting from a geomagnetic storm back to its origin on the Sun, which means that our simulated scenarios would not even fit in their studied cases. In the opposite direction, i.e. trying to predict the impact that CMEs would have on Earth, starting from remote sensing observations, these simulations could explain some issues in the current models. One example of such a study is the model of \citet{mateja_model_false_positives} which uses solar parameters to assess the occurrence probability of geomagnetic storms, which also contains false alarms, that is, CMEs that have the potential to impact Earth but did not produce any storm. Our simulations might be able to explain some of these false positive events, which should not be disregarded from a forecasting point of view, despite their lower-than-usual geoeffectiveness.\\                      

\begin{figure*}[h!]   
        \centering     
        \begin{tabular}{c c}
                {       \centering
                        \label{fig:Dst_Bz_Pdyn_SW_double_er}{\includegraphics[width=0.48\linewidth]{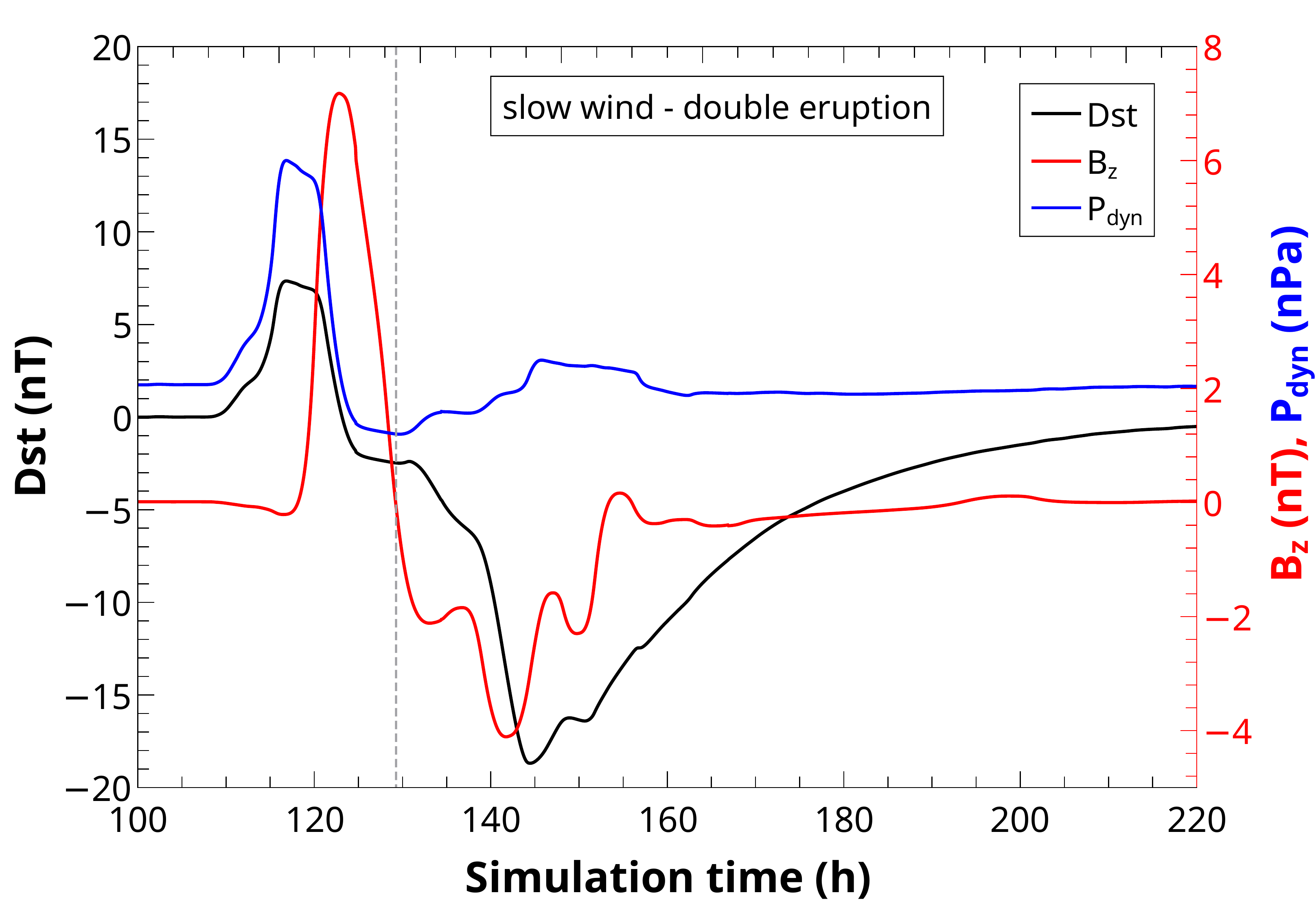}}                  
                }                      
                &
                {       \centering      
                        \label{fig:Dst_Bz_Pdyn_FW_stealth_speed}{\includegraphics[width=0.48\linewidth]{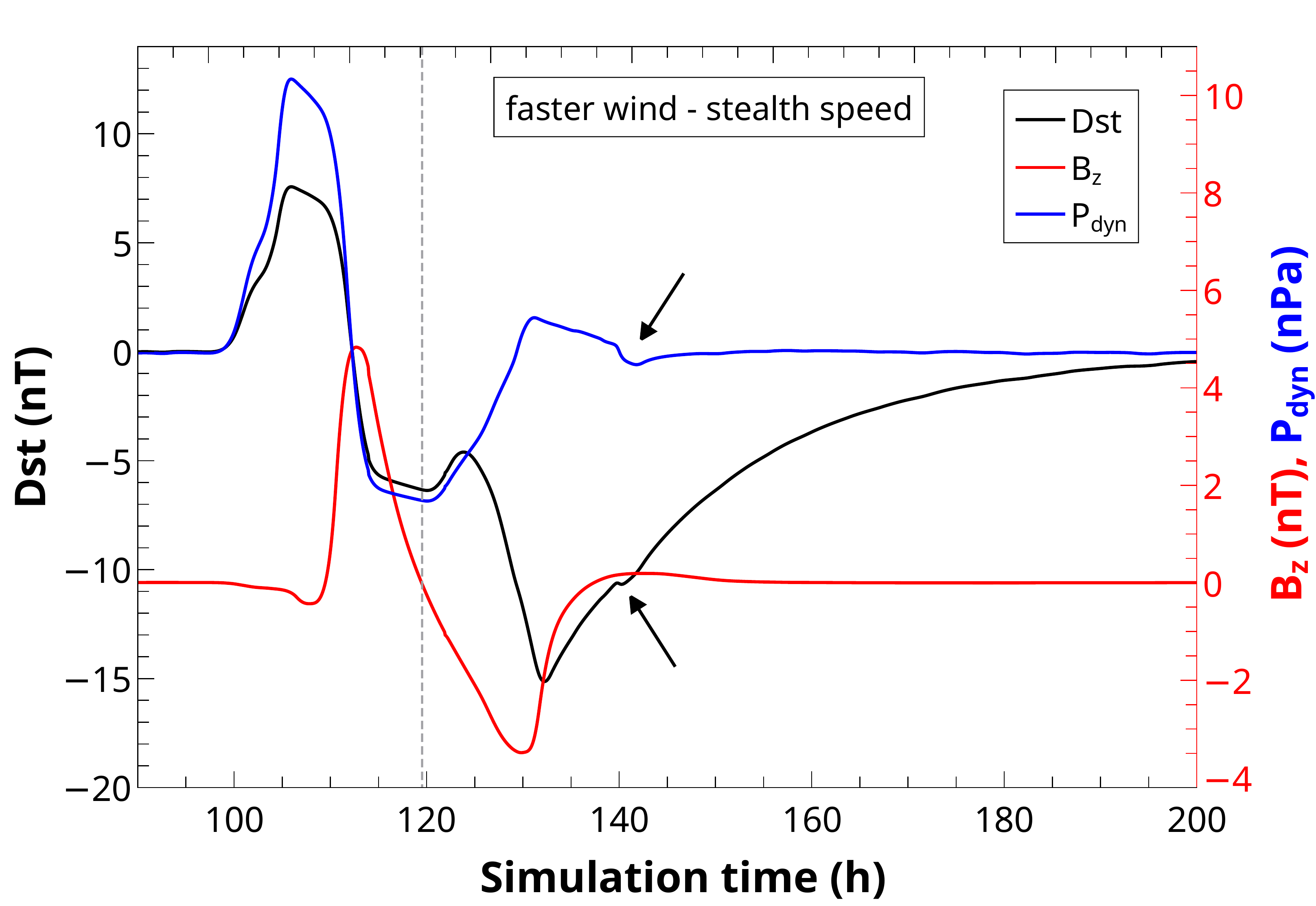}} 
                }                       
        \end{tabular}
        \caption{Modelled $Dst$ using simulation data (black line, left axis), the $B_z$ component of the magnetic field (red line, right axis), and the dynamic pressure (blue line, right axis) in the case of the slow wind - double eruption (left side) and faster wind - stealth speed (right side). The curves have not been overlapped, and so they indicate the actual simulated arrival time. The grey dashed line shows the polarity reversal of $B_z$.}
        \label{fig:Dst_Bz_Pdyn}
\end{figure*}

In order to analyse the different contributions of the solar wind parameters to the computed $Dst$, we plotted the evolution in time at 1 AU in the equatorial plane of the simulated $B_z$ magnetic field component, dynamic pressure, and $Dst$ in Fig.~\ref{fig:Dst_Bz_Pdyn}. The first feature that reaches Earth is a jump in pressure caused by an increase in speed in the first part of the ICME, which creates a compressed front. This small sudden storm commencement \citep[SSC; e.g.][]{SSC_mayaud_1975, SSC_curto_2007} can also be seen in the OMNI data in Fig.~\ref{fig:dst_omni_amrvac}, during the hours ahead of the green highlighted area. Before $B_z$ turns negative, the only contribution to the $Dst$ is given by the dynamic pressure, with the amplitude of $b\Delta P_{dyn}^{1/2}$, as also described initially by \citet{burton_dst_1975}. Once $B_z$ turns negative and the interplanetary electric field ($VB_S$) overcomes the 0.49 mV m$^{-1}$ threshold, the injection function $Q$ becomes negative and starts decreasing the $Dst$, creating the main phase of the storm. We note the oscillations in $Dst$ after its minimum value in the left panel of Fig.~\ref{fig:Dst_Bz_Pdyn}, which are imposed by the variations in $B_z$. On the other hand, in the right-hand plot, the decrease in $Dst$ just after SSC and increase during the main phase are both imposed by the profile of the dynamic pressure, which increases in response to plasma pile-ups. This contribution is indicated by the black arrows and can be more easily observed in the recovery phase of the storm, when $B_z$ is positive again and the energy injection from the solar wind into the magnetosphere has ended.

\begin{figure*}[h!]   
        \centering     
        \begin{tabular}{c c}
                {       \centering
                        \label{fig:dst_sims}{\includegraphics[width=0.48\linewidth]{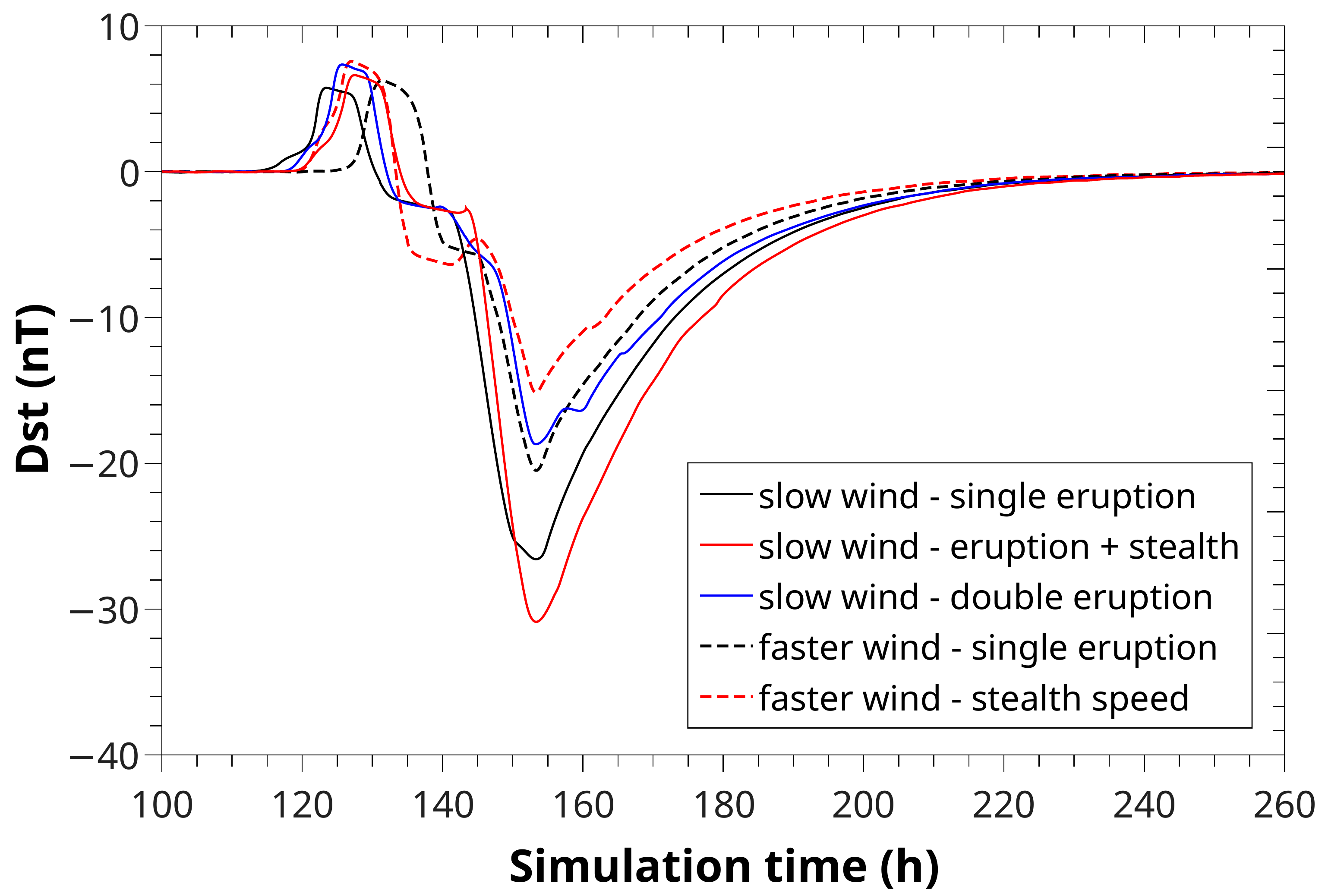}}                  
                }                      
                &
                {       \centering      
                    \label{fig:dst_sims_reversedBz}{\includegraphics[width=0.48\linewidth]{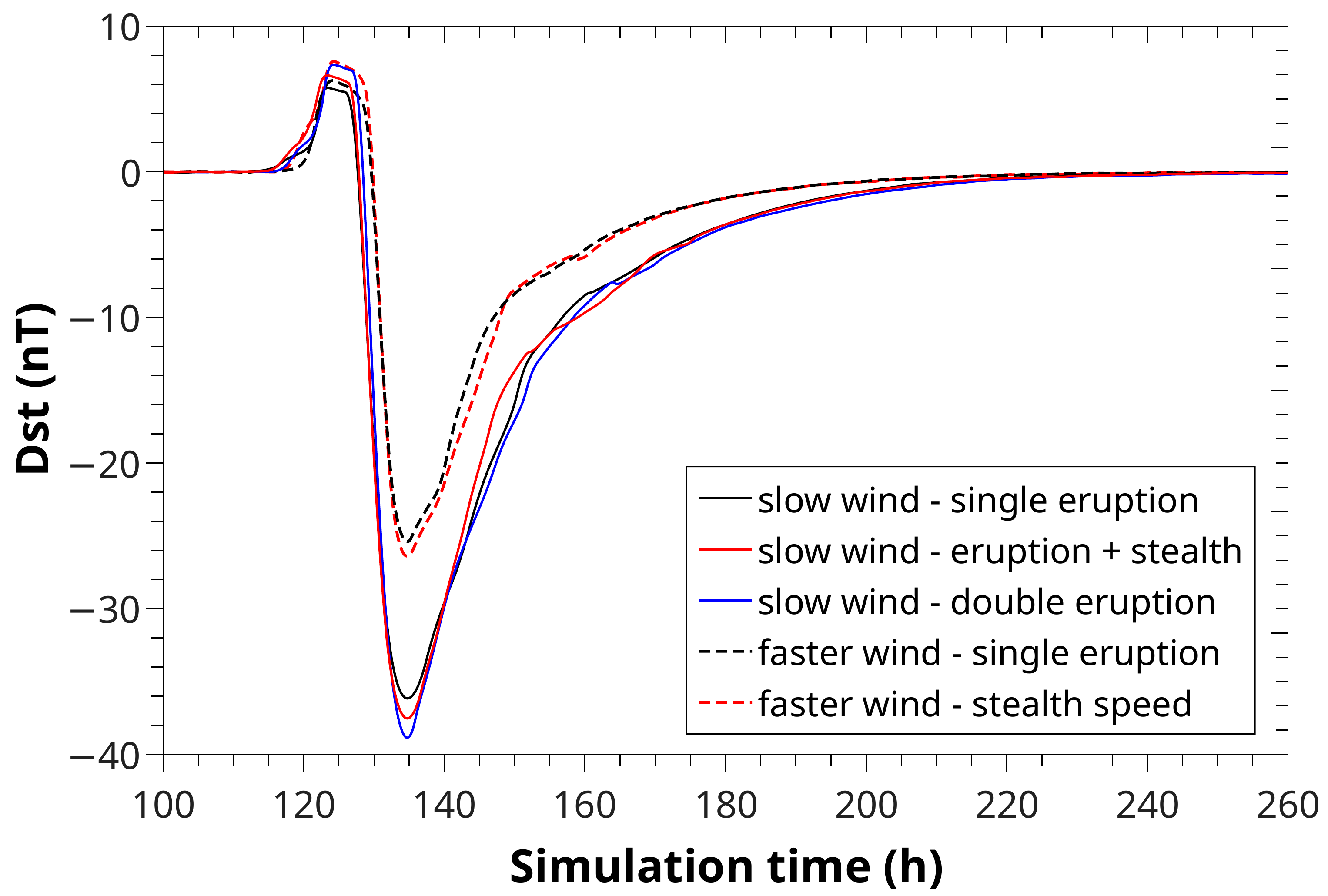}} 
                }                       
        \end{tabular}
        \caption{Modelled $Dst$ using simulation data (left side) and using the $B_z$ component with the reversed sign (right side). The minima of the curves have been aligned onto that of the last occurring minimum, which was the case of the single eruption in the slow wind.}
        \label{fig:Dst_simulations}
\end{figure*}

We computed and compared the $Dst$ indices for all the simulations, and their evolution is shown in the left panel of Fig.~\ref{fig:Dst_simulations}. The interplay between the two important aforementioned parameters slightly changes the order of the strength of the geomagnetic storms as compared to the $B_z$ minima plotted in Fig.~\ref{fig:sims_1AU_signatures}. The only difference is that the single eruption inserted into the faster wind becomes slightly more geoeffective than the double eruption in the slow wind. The least negative $B_z$ (faster wind, stealth speed case) still creates the weakest geoeffectiveness, whereas the most negative $B_z$ (slow wind, eruption + stealth case) induces a weak geomagnetic storm with the lowest $Dst$ of all the simulations.

Our proposed explanation for the overall low geoeffectiveness of the simulated eruptions is the reduction in the negative trailing $B_z$ component due to the magnetic reconnection with flux ropes appearing in the wake of the first CME. To test this hypothesis, we reversed the polarity of $B_z$ in the slices taken at 1 AU, such that the first CME would impact Earth with the negative front, and the reconnected part would be the positive one. We recomputed the $Dst$ using this new magnetic configuration, but keeping the same dynamic pressure and speed; the resulting values are plotted in the right panel of Fig.~\ref{fig:Dst_simulations}. As expected, in all simulations the frontal CME becomes more geoeffective, and all eruptions inserted into the slow wind produce weak geomagnetic storms, with an emphasis on the double eruption which now presents the lowest $Dst$ minimum. Interestingly, the strength of the storms is exactly correlated with the positive amplitude of the $B_z$ component in Fig.~\ref{fig:sims_1AU_signatures}, which would now become the negative front. Another aspect is the abrupt decrease in $Dst$ immediately after the SSC, meaning that the dynamic pressure did not exhibit a contribution as important as in the previous case, which is because the energy injection function turned negative as soon as the flux ropes arrived at Earth. This also led to a smaller variation in the index values between the simulations as well as during each individual storm. A final detail that can be extracted from Fig.~\ref{fig:Dst_simulations} is that even though the single eruption inserted into the faster wind was the only one that presented no following flux ropes and therefore lacked the trailing magnetic reconnection, it still changed geoeffectiveness along with the $B_z$ reversal of sign. Our understanding of these simulation results is that the frontal part of a CME has more impact on Earth's magnetosphere than the elongated tail and, as previously noted \citep[e.g.][]{fenrich_Bz_reversed_polarity}, this is exacerbated when it arrives with a negative N-S component of the magnetic field. 
                
 
    \section{Summary and Discussion} \label{sec:summary}  
In this paper, we describe and analyse five simulations performed using the code MPI-AMRVAC, of which three were propagated into a slow wind, and two were inserted into a faster and denser background wind. The model configuration is based on an MCME event on 21-22 Sept 2009 that erupted approximately towards Mercury and Earth, providing an opportunity to compare the in situ data between simulations and two spacecraft. Initial simulations of this event were presented in paper I, where good agreement was found between coronal remote sensing observations and the double eruption and eruption + stealth scenarios ejected into the slow wind. In this paper, we propagate these two cases out to 1 AU. At 0.3 AU, the eruption with the stealth ejecta shows the best fit to in situ data taken from MESSENGER, which is surprising when compared to the results of paper I where we found a better fit of white-light coronagraph observations with the double eruption scenario. This can be attributed to our 2.5D setup, as the slices are not affected by the longitudinal difference between CME propagation and spacecraft, and therefore in the MESSENGER data some flux ropes might be missed due to their narrow dimensions. The stealth ejecta is also partially missed in simulations, which is attributed to the slice being taken at the equator, while the current sheet in which the CME propagates is slightly shifted northward. At Earth, the differences between the two simulations are even smaller because of the influence of the solar wind throughout the propagation. Taking all these factors into consideration, it is difficult to distinguish the triggering mechanism responsible for the second observed CME, making the distinction between stealth ejecta and source CMEs even more unclear. This highlights the need for better remote sensing instruments with higher resolution and cadence, which may in the future be able to observe much fainter structures, allowing us to better understand these stealth CMEs.\\ 
We also analyse the influence of the solar wind on the eruptions by numerically simulating two more scenarios of shear-induced CMEs in a faster and denser background wind. A first observation we extracted from this new configuration was a change in the initial magnetic structure, even though we have not interfered with the magnetic components, leading to an opening of the overall streamer. Even if our magnetic solution is not originating from a potential field, this could still indicate a non-unique solution for extrapolations that compute the coronal magnetic field solely from line-of-sight magnetic field converted to $B_r$, as in the potential field source surface \citep[PFSS,][]{pfss1969} model. The PFSS solution also depends on both lower and upper boundary conditions of the other non-radial components of the magnetic field, and according to those, one can obtain different resulting magnetic configurations. In order to better reproduce the field lines in the solar atmosphere, one might need to better constrain the plasma solution by providing the other $B$ components, or other parameters such as density and/or temperature from observations. \citet{judge_inversion} described a method of inversion of spectropolarimetric data from observed coronal plasma at the limb of the Sun in order to obtain the magnetic field components inferred from measured Stokes parameters $I$, $Q$, $U,$ and $V$. Such a method can be used to invert physical parameters from coronal observations taken by the novel Daniel K. Inouye Solar Telescope \citep[DKIST,][]{dkist2}, and an inversion algorithm is currently being prepared by \citet{paraschiv}.\\ 
The change in our simulated magnetic field configuration also affected the subsequent eruptions, by creating two CMEs from the applied shear when using the $v_0$ value that resulted in a stealth ejecta in the slow wind case. There were also no plasma blobs appearing in the faster wind simulations, making them a possible indicator of the initial magnetic structure.\\
In addition, we computed the $Dst$ index from our simulations using an empirical formula into which we introduced the speed, the $B_z$ component of the magnetic field, and the dynamic pressure. The measured $Dst$ is both qualitatively and quantitatively reproduced by the double eruption scenario in the slow wind, as is the small SSC occurring before the drop in $Dst$. The computed recovery phase is longer and more gradual than the observed one, which we can attribute to another event developing after the passage of our ICME that  again decreases the actual $Dst$. That second event was only identified in in situ data and is not investigated in the presented study because it is of no current interest. We also analysed the different parameters contributing to the computed $Dst$ and concluded that the amplitude of a storm can be greatly influenced not only by $B_z$ but also by the dynamic pressure. The low geoeffectiveness of the slow wind simulations was attributed to the magnetic reconnection in the tail of the CMEs, which made the trailing $B_z$ less negative, leading us to study the $Dst$ in the case where the CMEs arrived at Earth with reversed polarity of $B_z$. This change in the polarity of $B_z$ increased the geoeffectiveness of all the simulated eruptions and decreased the contribution of the dynamic pressure to the overall trend of the storms, which now affected mainly the SSCs. Another finding from this study is that multiple CMEs might even have a reduced geoeffectiveness as compared to single CMEs, which could explain some false alarms encountered in storm predictions using solar parameters. In a follow-up paper we will present the analysis of forces that contribute to the eruption of these CMEs, mergers between them, and their propagation to 1 AU.

    \begin{acknowledgements}
    We thank the referee for their constructive comments and corrections, which led to the improvement of this manuscript. The authors are grateful to Emmanuel Chan\'{e} and to Ilia Roussev for all the insightful discussions and thank Luciano Rodriguez for the help in the interpretation of in situ signatures. We also thank Ian G. Richardson for the thorough read of the paper and for all the valuable comments and suggestions. D.C.T. was funded by the Ph.D. fellowship of the Research Foundation – Flanders (FWO), contract number T1 1118918N. D.C.T., E.D. and M.M. acknowledge support from the Belgian Federal Science Policy Office (BELSPO) in the framework of the ESA-PRODEX program, grant No. 4000120800. This work was partially supported by a PhD Grant awarded by the Royal Observatory of Belgium. These results were also obtained in the framework of the projects C14/19/089 (C1 project Internal Funds KU Leuven), G.0D07.19N (FWO-Vlaanderen), SIDC Data Exploitation (ESA Prodex-12), and BELSPO projects BR/165/A2/CCSOM and B2/191/P1/SWiM. For the computations we used the infrastructure of the VSC – Flemish Supercomputer Center, funded by the Hercules foundation and the Flemish Government – department EWI.
    We acknowledge use of NASA/GSFC's Space Physics Data Facility's OMNIWeb (or CDAWeb or ftp) service, and OMNI data. We acknowledge the use of STEREO/SECCHI data (http://stereo-ssc.nascom.nasa.gov/). The $Dst$ index used in this paper was provided by the WDC for Geomagnetism, Kyoto.

    \end{acknowledgements}

    \bibliographystyle{aa}
    \bibliography{talpeanu.bib}
        
\end{document}